# Optimization of airgap in a monocentric lens assembly and metasurface based anti-reflecting coating in the long-wave IR regime

Manish Kala[1,†], Pawan Singh[1,2,†], Sanjay Kumar Mishra[3], Unnikrishnan Gopinathan[3], Ajay Kumar[3], Akhilesh Kumar Mishra[1, 4, *]

[1]*Department of Physics, Indian Institute of Technology Roorkee, Roorkee-247667, Uttarakhand, India*

[2]*Department of Physics, National Institute of Technology Warangal, Warangal-506004, India*

[3]*Photonics Research Lab, Instruments Research and Development Establishment, Dehradun 248008, India*

[4]*Centre for Photonics and Quantum Communication Technology, Indian Institute of Technology Roorkee, Roorkee-247667, Uttarakhand, India*

[†] *These authors contributed equally to this work.*

[*]*Corresponding author: akhilesh.mishra@ph.iitr.ac.in*

**Abstract**

Owing to minimal aberration, larger field-of-view, and high resolution, monocentric lenses are preferred over other imagers in hemispherical image surface scenarios, particularly in the long-wave infrared (LWIR) region. Herein, we study a monocentric lens assembly consisting of a ball lens and two hemispherical shell lenses of given radii of curvature and lens materials to enhance the transmission of LWIR ranging from 8μm to 12μm wavelength. We optimize the air gaps between the lenses of the assembly for a wider range of incidence angles, assuming antireflecting coating (ARC) on all surfaces of the lenses. Additionally, the effect of temperature variation on the transmittance is also studied to attain an optimum air gap between the lenses. With the variation in incidence angle and temperature, modulations in focal length and spot size are reported for different wavelengths. Further, we propose a metasurface design to replace conventional multilayer ARCs to enhance the transmittance through the lens assembly over the given wavelength range for a wide field-of-view.

## 1. Introduction

The imaging technology has witnessed a rapid, continuous growth over the last few decades. In particular, applications such as in aircraft monitoring, surveillance, communications, agriculture, forensics, etc., demand a wide field-of-view (FOV) along with higher-resolution images [1-6]. To achieve higher resolution, it is required to match the resolution of the objective to that of the detector. This in turn maximizes the information of the imaging system quantified by space bandwidth product (SBP). Rapid advancement in detector technology poses challenges of designing optics with resolving powers that match detector resolution. One of the strategies to increase the resolution and thereby SBP for a given FOV is to scale the optical system (objective and the detector) [7].

For high-resolution images, the lens assembly must have a large numerical aperture to gather radiation efficiently and a long focal length to magnify the image with the least aberrations [8]. To overcome the problem, a wide-field objective lens is associated with cameras with multiple narrow-field micro cameras, which reduces aberration in the image. The large aperture of a micro camera can resolve the problem of aberration, but requires a complex optical system. But, the small aperture size of the micro camera captures less incoming radiation [9-11]. For a wider FOV, a monocentric lens assembly, which is a combination of a spherical lens (ball lens) and multiple spherical or hemispherical shells, is used which does not face coma and astigmatism [12]. In a monocentric lens, light can be controlled by a stop aperture at the center of the lens or by transfer optics and the image is supposed to be formed on the curved surface but due to the complex design of a curved detector or the unavailability of the detector, a large number of flat surfaces are used to mimick the curved image surface. Alternatively, it can be done by imaging the fiber bundle with a curved input and a flat output surface [19]. Stamenov et al. demonstrated angle-independent resolution for conventional imagers for a broad FOV, and they compared imaging and waveguide-based transmission images through monocentric lenses [20]. For achieving a wider FOV with spherical surfaces, an early design of the monocentric lens was introduced by T. Sutton in 1859, but it had uncorrected aberrations and a high *f*-number. To resolve these problems, in 1942, J. G. Baker suggested to use two different glasses with high (flint glass for shells) and low refractive indices (crown glass for core lens), which offered the correction in aberration and *f*-number [13]. Mostly, the two-glass system (2GS) monocentric lens works well for a wide FOV. The 2GS-based monocentric lens assembly displays a cutting-edge innovation in optical and thermal imaging that enhances visual performance by the integration of two IR glasses within a single assembly for versatile applications [19-22]. In the multiscale imaging system, the use of a monocentric lens assembly as an objective lens was discussed by Ford and Trembly [14]. In such a system, overlying parts of the spherical image surface are imparted onto image detectors or sensors, which form the whole image by the combination of all images detected by multiple sub-imagers [14,15]. The FOV of sub-imagers must overlap efficiently for image formation [17,18]. To minimize the aberration, Uniyal et al. proposed a multiscale monocentric lens imager using first-order optics for the LWIR region, which covers $80^{o}\times16^{o}$ FOV over five detector arrays [23].

The monocentric lens being a critical component in the development of wide FOV, and high-resolution imaging systems, its design, fabrication, and assembly are of utmost importance. In general, transmission-related issues are seldom considered in lens design because suitable anti-reflection coatings are widely available, and no significant air gaps are typically present between lens elements. But, in a monocentric lens, the radius of curvature of the inner surface of the core lens and the inner surface of the shell lens must be identical, which is extremely difficult to fabricate with high precision. In addition, a major issue is the unavailability of the suitable optical glues to assemble lenses in the IR domain. These make the realization of such a monocentric lens assembly difficult, as leaving an airgap between the two lenses becomes indispensable. This paper seeks to address this research gap. Since the refractive indices of air and lens material are very different, the associated transmittance losses are huge. Hence, the airgap optimization is essential to optimize transmission in such systems. Herein, we optimize the airgaps between the lenses in a two-glass monocentric lens assembly of given radii of curvature and lens materials. The lenses of the assembly are assumed to be coated with ARCs to achieve maximum transmittance in the long-wave infrared (LWIR) region ($8\ \mu m$ to $12\ \mu m$ wavelength) encounters the problem of low transmittance and distortion due to large refractive index contrast. To enhance the transmittance, the air-gaps between ARCs are primarily optimized to prevent the collapse of material stability with temperature variations. In this work, we consider a monocentric lens with IG6 glass as a core lens and two hemispherical shells made up of Ge for LWIR radiation for a wide FOV. The effect of temperature variation is also discussed, which further optimizes the airgap and output transmittance. Additionally, we propose a robust, thermally stable metasurface design on the lens surfaces, holding the potential of replacing conventional ARCs.

## 2. Monocentric lens configuration

Monocentric lens assembly consists of one spherical core lens (also called ball lens) and two concentric hemispherical shells as shown schematically in Fig. 1. We have considered the core lens ($L_2$) of material IG6 and hemispherical shells ($L_1$ & $L_3$) of Ge. The spherical core lens $L_2$ is sandwiched between two hemispherical surfaces or lenses $L_1$ and $L_3$ as shown schematically in Fig.1a. The lens assembly focuses the incident rays on a spherical plane placed after lens $L_3$ at the focal length of the lens assembly (see Fig. 1b).

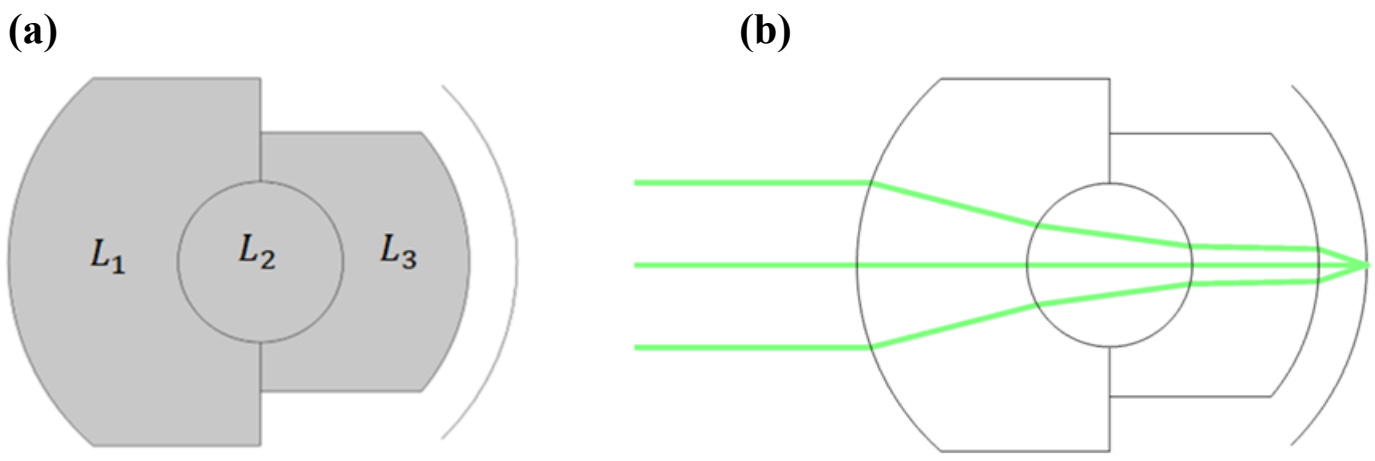

**Fig. 1**: (a) Schematic of the monocentric lens assembly, (b) focusing of rays on a spherical image plane. In lens assembly, $L_2$ represents the ball lens (IG6), and $L_1$ & $L_3$ are hemispherical lenses (Ge).

As shown in Fig. 2, we have also considered an air gap between core lens $L_2$ and two internal surfaces of hemispherical shells ($L_1$ & $L_3$). Both internal and external surfaces of all the lenses are assumed to be coated with anti-reflecting coatings (ARCs).

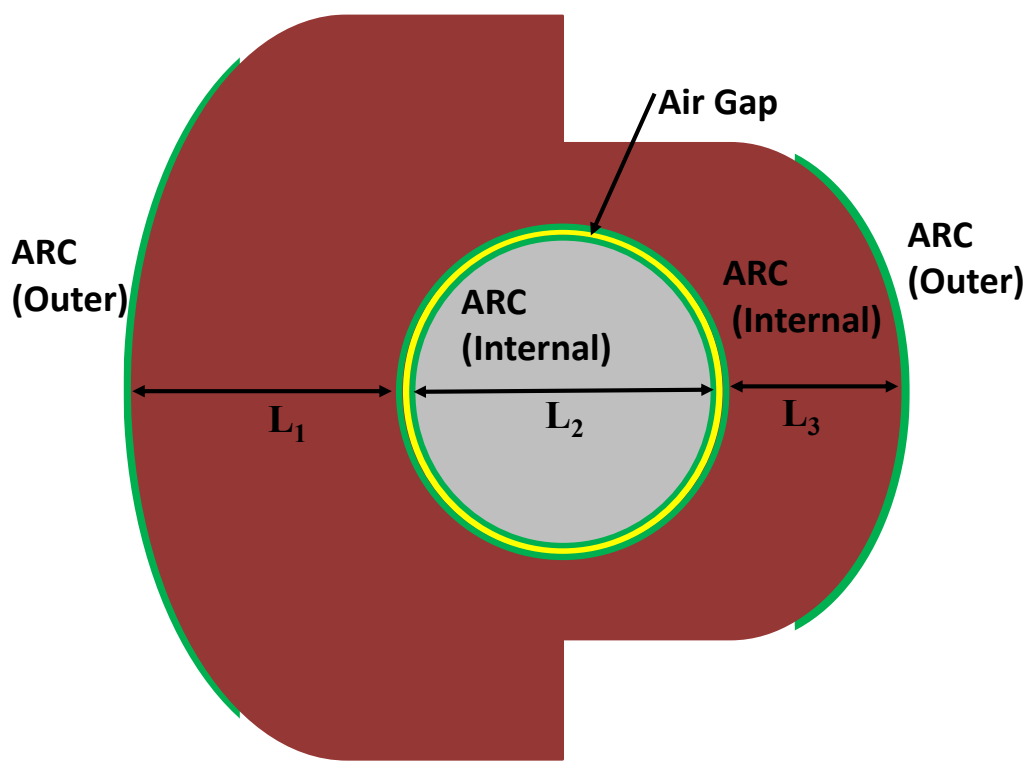

**Fig. 2**: Representation of monocentric lens assembly with an airgap (yellow) and with internal and outer ARCs (green), lenses $L_1$ & $L_3$ are in magenta and while $L_2$ is represented by gray colour.

The combination of the three lenses covers a wide FOV for LWIR radiation (8 μm – 12 μm). In our calculation, we have considered two hemispherical shells $L_1$ and $L_3$ with radii of curvature (ROC) 88.75 mm and 73.05 mm, respectively, while the core lens $L_2$ has ROC of 28.35 mm. The lens assembly parameters are listed in Table 1.

**Table 1**: The parameters of the monocentric lens assembly

| Lens | ROC (mm) | Diameter (mm) | Material |
|---|---|---|---|
| $L_1$ | $R_1$ = 88.75 (left)<br>$R_2$ = 28.35 (right) | 150 | Ge |
| $L_2$ | $R_1$ = 28.35 (left)<br>$R_2$ = 28.35 (right) | 56.7 | IG6 |
| $L_3$ | $R_1$ = 28.35 (left)<br>$R_2$ = 73.05 (right) | 108.3 | Ge |

The refractive indices of Ge and IG6 are taken from refs. [24, 25]. ARC thickness on the surfaces of lenses is assumed to be 1.25 μm.

We calculate the transmittance of the monocentric lens assembly using COMSOL® Multiphysics software [26]. In the design of lens assembly, the diameter of the aperture or entrance pupil is kept to be 52 mm and for overlapping angular positions, the entrance pupil size is kept fixed at 91.7 mm. For studying transmittance with airgap variation, all the interfaces are assumed to have ARCs with a transmittance of 97%. Initially, an airgap is considered to be 10 μm between internal ARCs of the IG6 ball lens and Ge hemispherical lenses, and we consider the IR radiation with a wavelength ranging from 8 μm to 12 μm for the wide incident angles. Moreover, we have calculated the transmittance with two distinct designs of metasurface on IG6 and Ge lenses replacing traditional multilayer ARCs.

## 3. Results and discussion

The optimization study of monocentric lens assembly is divided into three subsections. In the first subsection, we study the transmittance of IR radiation through a lens assembly with a varying air-gap between internal ARCs. In the second subsection, we highlight the effect of temperature variation on the transmission of IR radiation, focal length, and spot size for various incident angles from the input grid. Two different metasurface designs proposed to replace the conventional ARCs are detailed in the third sub-section.

To perform the simulation studies, we have used the finite element method-based COMSOL® Multiphysics platform and optimized 3D geometry (see Fig. 3a). The discretized geometry of lenses is shown in Fig. 3b. For the release of rays on lens assembly, we have considered a circular grid of diameter of 52 mm such that the center of grid and monocentric lens are on the same axis. The circular grid from the source side and the detector side are shown in Fig. 4 (a) and (b), respectively. The focusing of rays through the lens assembly is shown in Fig. 5, which shows all the rays are focused on the center of the image plane at a focal length 91.7 mm.

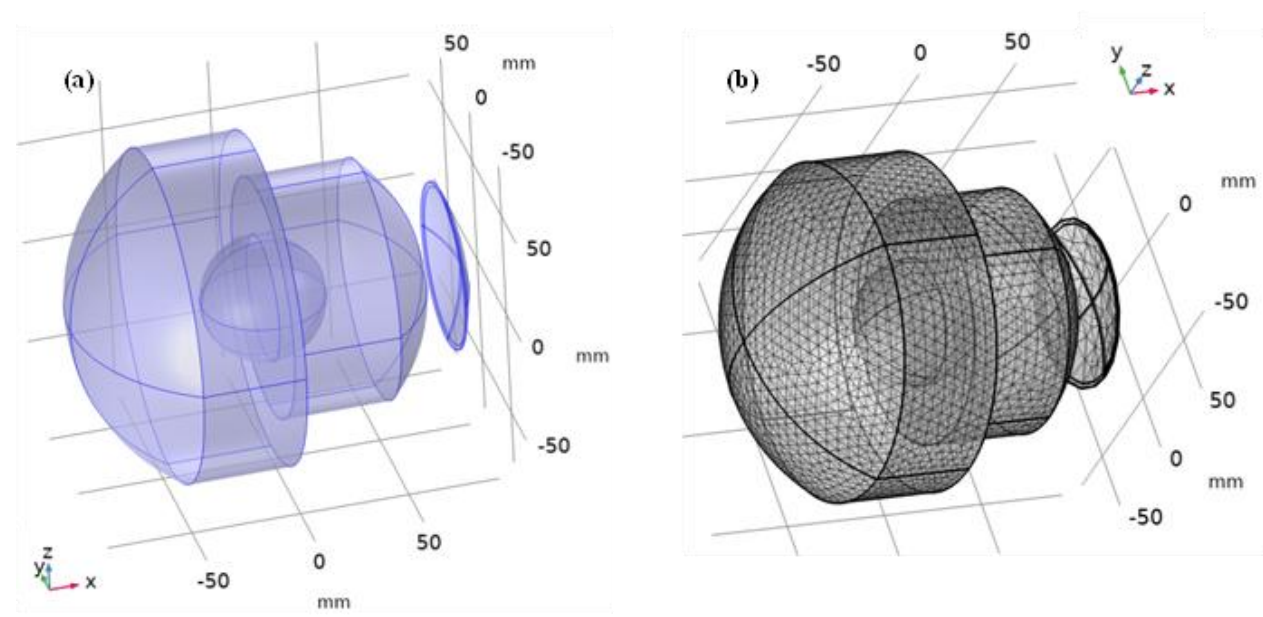

**Fig. 3:** (a) 3D geometry and (b) meshing of monocentric lens.

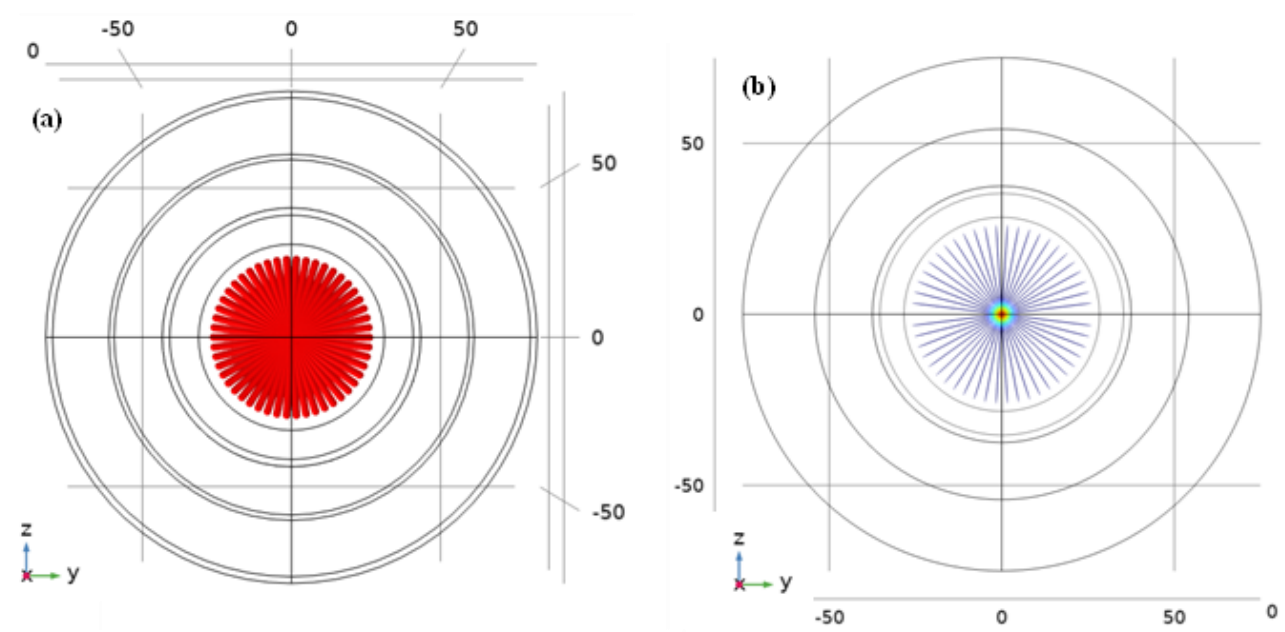

**Fig. 4:** Circular grid (a) source side to release the rays (b) detector side to receive the rays.

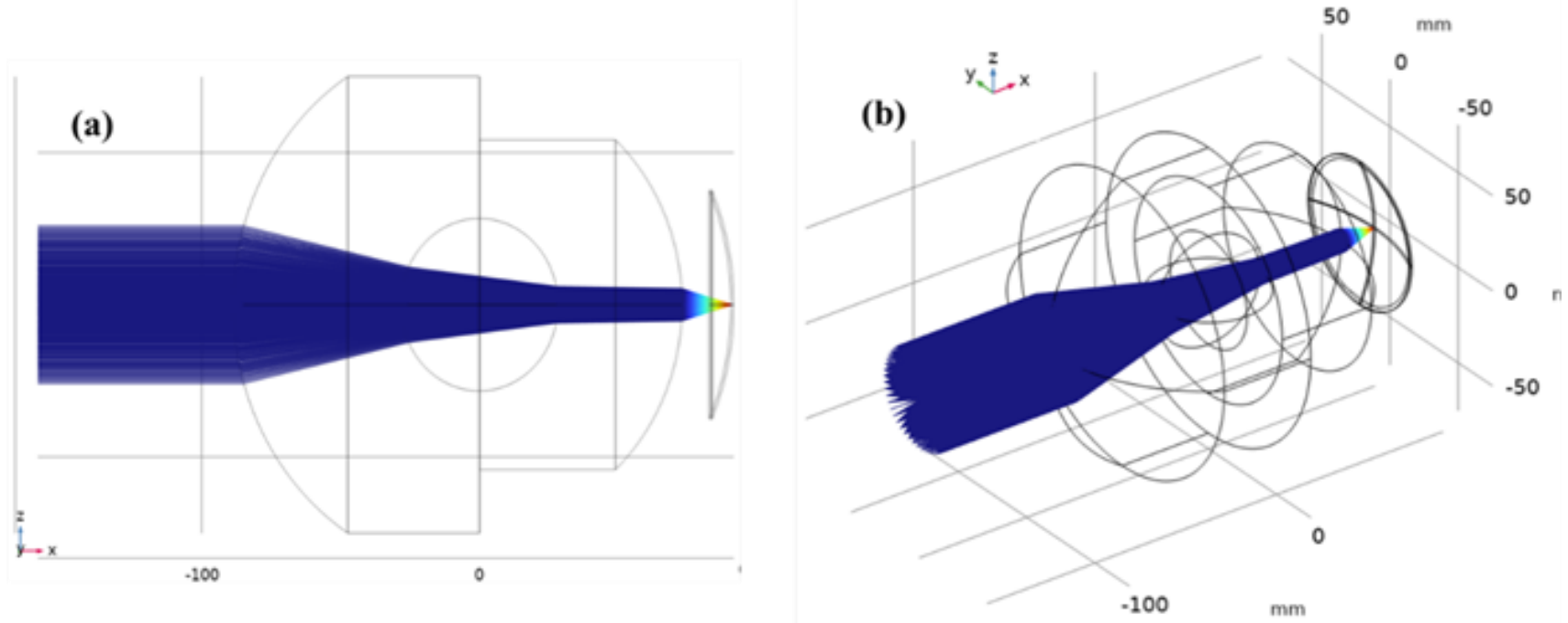

**Fig. 5:** Focusing of rays on the curved screen behind the lens at 91.7 mm distance from center (a) 2D view (b) 3D view.

### *3.1 Variation of the transmittance of the IR lens with airgap*

For optimizing transmittance, we have calculated the transmittance with airgap variation as shown in Fig. 6 at normal incidence. At 10 μm wavelength, the transmittance maximizes to 58% for 2.5 μm airgap, and further increase in airgap leads to oscillation in transmission. We noticed that the average (over wavelength) transmittance almost remains constant (maximum 53.7%) for the wavelength range $8 - 12$ μm.

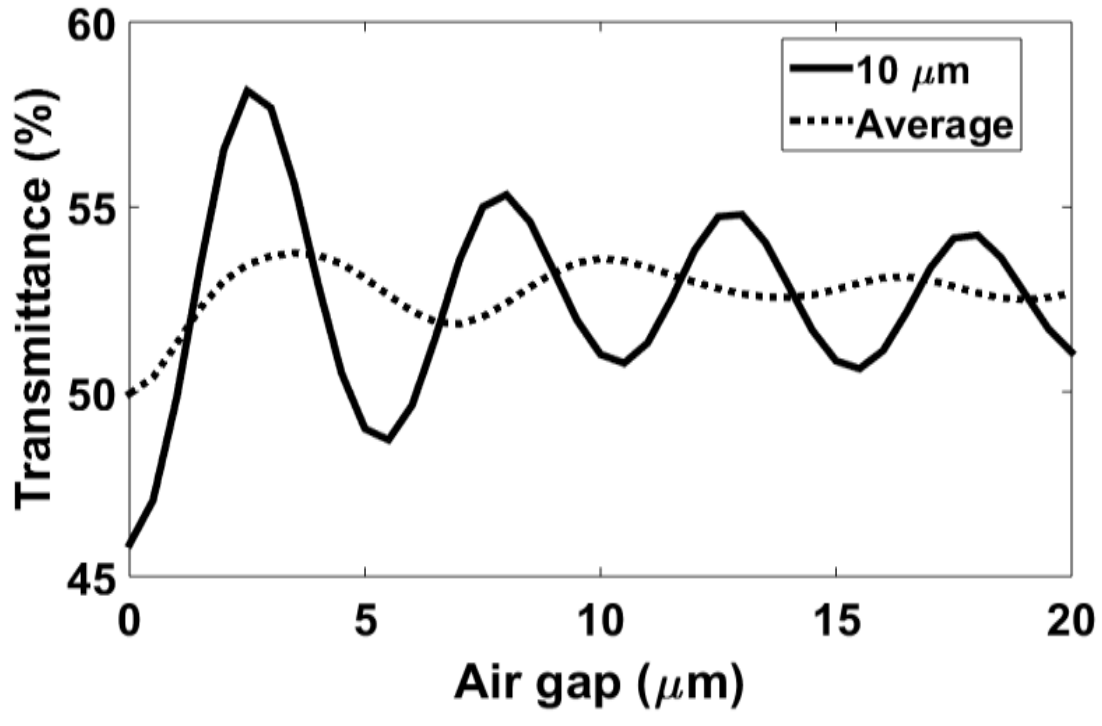

**Fig. 6:** Transmittance of 10μm wavelength, and average transmittance over all wavelengths ($8 -$ 12 μm) with airgap for normal incidence. Solid and dashed lines show the transmittance for 10 μm wavelength and average transmittance over the wavelength range $8 - 12$ μm, respectively.

Next, we study the effect of the variation of the angle of incidence on transmittance. The average transmittance over all wavelengths ($8-12$ μm) at different angles of incidence (0°, 2°, 4°, 6°, and 8°) is shown in Fig. 7 (a). From the figure, it is clear that the average transmittance decreases with increasing angles of incidence. At normal incidence, the lens assembly shows about 53.7% average transmission. A comparative study of average transmittance over all angles of incidence for 10 μm and averaged over $8-12$ μm wavelengths are shown in Fig. 7b. The figure explains that average transmittance at 10 μm wavelength averaged over all angles of incidence is 52.7 % at 2.5 μm airgap, while the wavelength average transmittance over all angles for IR radiation ($8-12$ μm) is only 48.3% at 2 μm airgap. The variation in spot size and transmittance for the lens assembly is given in supplementary [27].

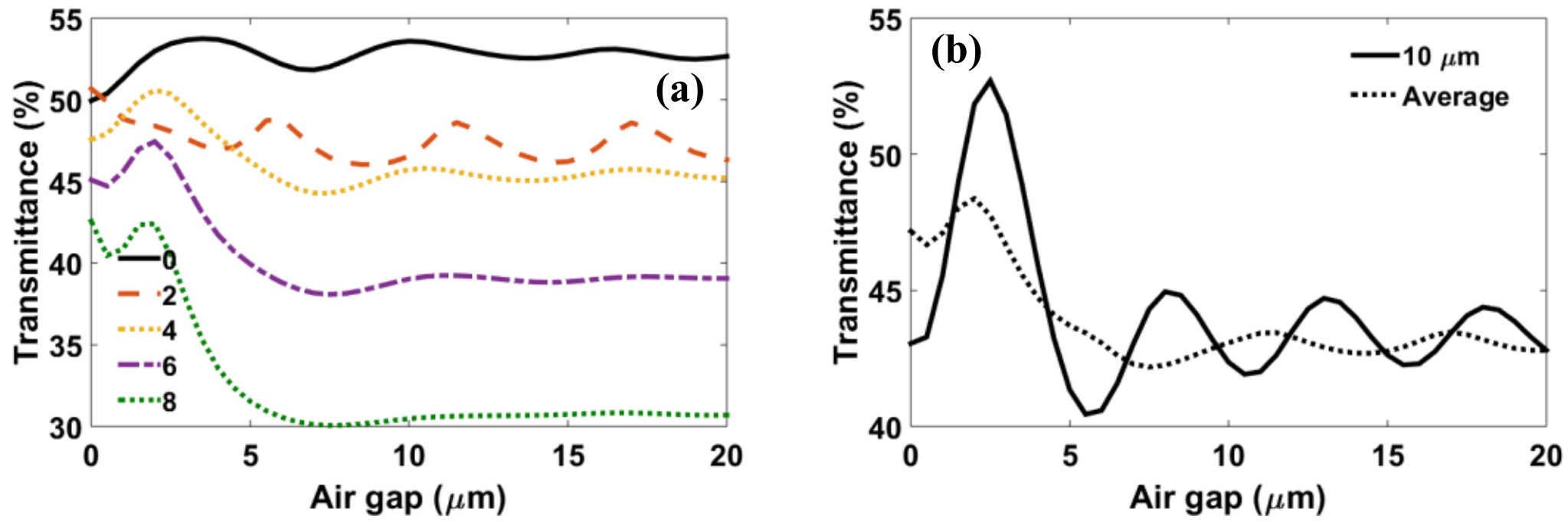

**Fig. 7:** (a) Variation of average transmittance over all wavelengths ($8-12$ μm) with airgap at different incidence angles and (b) solid and dashed lines show the angle-averaged transmittance for 10 μm wavelength and wavelength-averaged transmittance over wavelength range $8-12$ μm, respectively.

As an extension of this study, we also examine the effect of temperature variation (ranging from 0°C to 40°C) on the transmittance of the lens assembly. As the temperature increases, the refractive indices and the radii of ball and hemispherical lenses vary, which in turn alters the airgap between the lenses. Hence, the temperature variation leads to variation in the transmittance. In addition, temperature variation also changes focal length and spot size, which are detailed in the supplementary [27].

### ***3.2 Effect of temperature on transmittance of IR radiation***

To study the temperature dependence of the transmittance, we have considered the wavelength as well as temperature dependence of the refractive indices of IG6 and Ge, while ARCs are assumed to be independent of the temperature variation. Initially, we assumed 10 μm airgap, which of course varies with the temperature. With an increase in the temperature, the refractive indices of Ge and IG6 increase linearly, while it decreases with an increase in the wavelength [24, 25].

In the lens assembly, temperature not only affects the refractive indices but also alters the thicknesses of lenses. The expansion of lens thickness $(d)$ can be studied using $d = d_0 + d_0\alpha(T_m - T_0)$, which shows that the thickness of the lens increases linearly, where $T_m$, $T_0$, and $\alpha$ are operating temperature, reference temperature $(20^o C)$ and the coefficient of thermal expansion of the material, respectively [28]**.** The expansions of Ge and IG6 with temperature are shown in Figure 8 (a). On increasing the temperature, the outer surface of IG6 grows more rapidly than the internal surface of Ge lens. As both materials expand, the airgap between ARCs decreases gradually (see Fig. 8 (b)). Therefore, the optimization of the airgap is crucial to prevent damage of ARCs due to lens contact and the mechanical stability of the assembly.

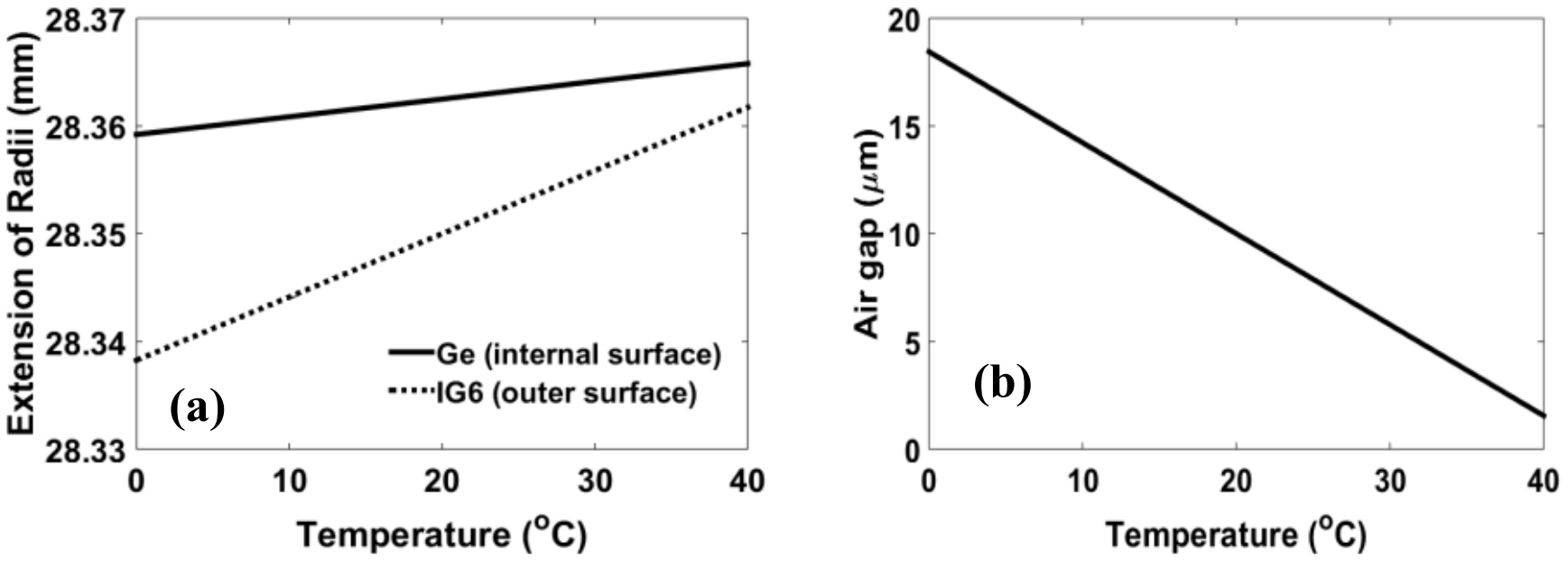

**Fig. 8**: (a) Thermal expansion of Ge (internal surface) and IG6 (outer surface), and (b) reduction of the airgap (initial gap 10 μm) with temperature.

To avoid the collapse of internal ARCs, we have considered the airgap to be 10 μm initially. Fig 8 (b) shows that reduction of the airgap with temperature up to $40^{o}C$, where the airgap has reduced to 1.55 μm. Also, if we reduce the temperature, then the airgap increases due to the contraction of the lens materials, and at $0^{o}C$, we have 18 μm airgap between the two lens surfaces as depicted in Fig. 8 (b).

On comparing the average transmittance of lens assembly over the wavelength range $8\ -12\ \mu m$ as shown in Fig. 9 (a), we note that as the incident angle increases, the average transmittance decreases down to 24% (for 8° incident angle) and it maximizes to 49.5% for normal incidence. If we take the average over all wavelengths and all incident angles, the transmittance is found 37%, but it exhibits a sudden increase if we increase the temperature beyond 30°C due to a larger effective optical path length for a wider incidence angle. At 40°C temperature, the transmittance is calculated to be ~ 45% due to reduced airgap, as shown in Fig. 9 (b).

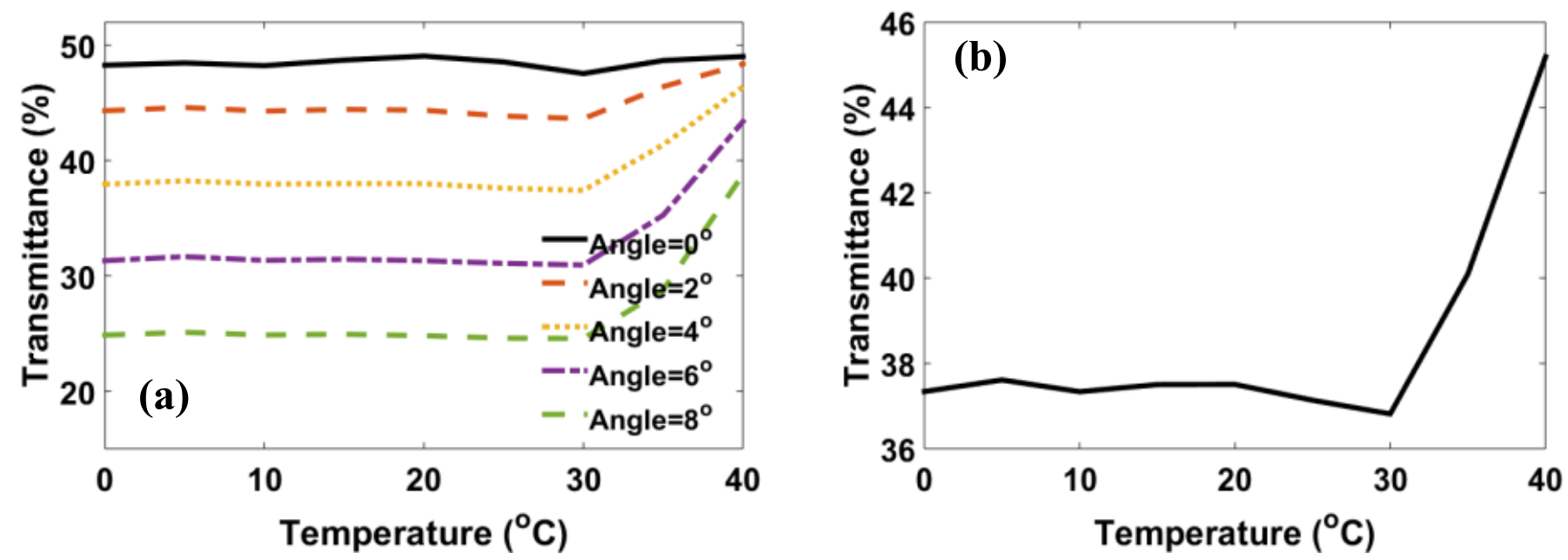

**Fig. 9**: (a) Averaged transmittance over all wavelengths for different angle, and (b) average transmittance over all wavelengths and all angles of incidence with increasing temperature.

Apart from affecting the transmittance, the temperature also alters the size of the focal spot and focal length, which are tabulated in Tables 2 and 3. From the tables, it can be observed that the focal length increases with temperature, but it decreases with incident angles, whereas spot size increases with both temperature and incidence angle.

A representative focal spot and focal plane at 10 μm wavelength and 20°C temperature for normal incidence is shown in Fig. 10. The relevant parameters are also given in the figures.

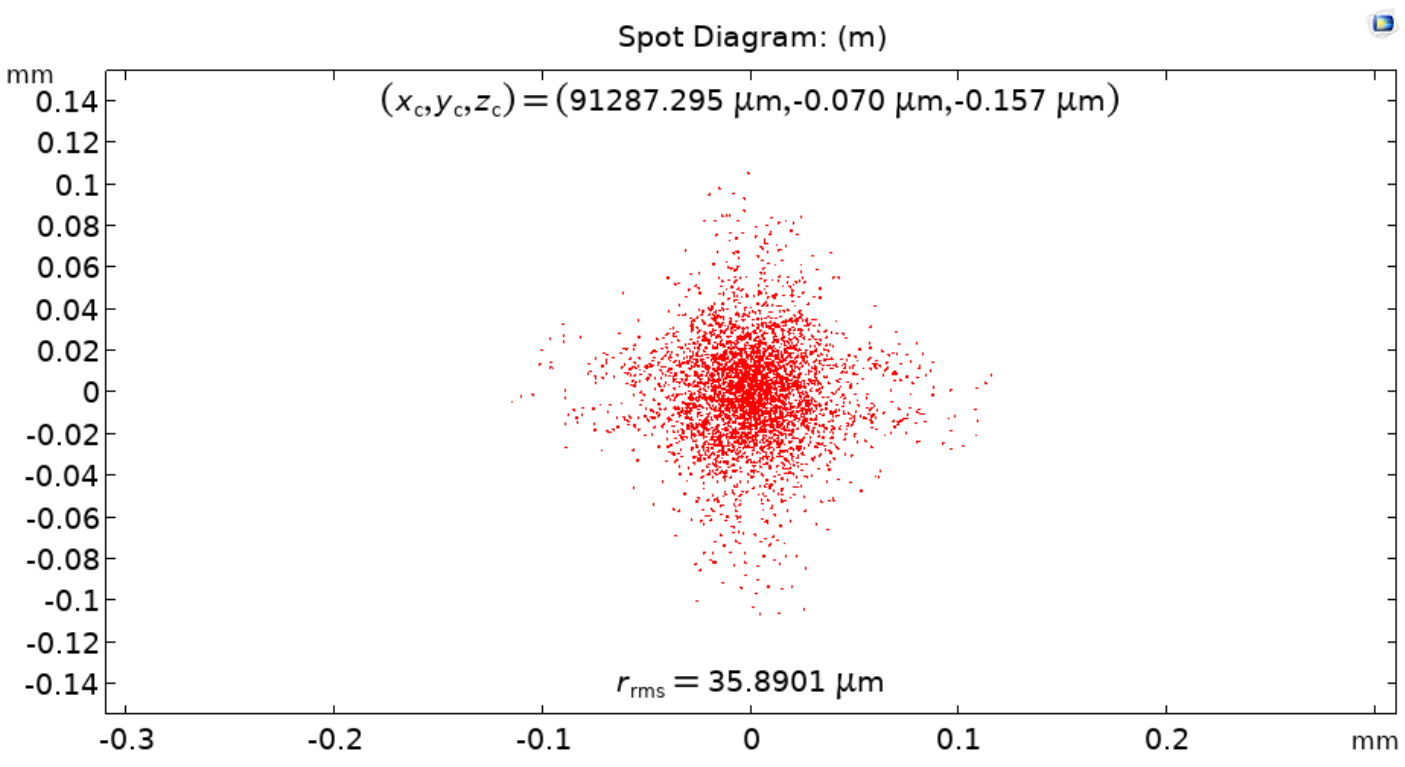

**Fig. 10:** Focal spot size at $10\mu m$ wavelength at 20°C for normal incidence.

**Table 2:** RMS value of focus spot size on image surface at 10 $\mu m$ wavelength for different incident angles

| | | **Calculated RMS value of Focus spot (µm) at 10µm wavelength** | | |
|---|---|---|---|---|
| **S. No.** | **Temperature (°C)** | **Angle of incidence=0°** | **Angle of incidence=4°** | **Angle of incidence=8°** |
| 1 | 0 | 35.89 | 35.76 | 48.23 |
| 2 | 5 | 35.89 | 35.88 | 48.04 |
| 3 | 10 | 35.86 | 35.73 | 47.90 |
| 4 | 15 | 36.05 | 35.83 | 47.74 |
| 5 | 20 | 36.00 | 35.78 | 47.57 |
| 6 | 25 | 36.14 | 35.82 | 47.57 |
| 7 | 30 | 36.31 | 35.75 | 47.37 |
| 8 | 35 | 36.14 | 35.81 | 47.06 |
| 9 | 40 | 36.33 | 35.77 | 47.05 |

**Table 3:** Focal length of lens assembly at 10 µm wavelength for different incident angles

| | | **Calculated focal length (mm) at 10 µm wavelength** | | |
|---|---|---|---|---|
| **S. No.** | **Temperature (°C)** | **Angle of incidence=0°** | **Angle of incidence=4°** | **Angle of incidence=8°** |
| 1 | 0 | 91.28 | 91.08 | 90.57 |
| 2 | 5 | 91.31 | 91.15 | 90.60 |
| 3 | 10 | 91.34 | 91.17 | 90.63 |
| 4 | 15 | 91.37 | 91.21 | 90.66 |
| 5 | 20 | 91.40 | 91.24 | 90.69 |
| 6 | 25 | 91.44 | 91.28 | 90.73 |
| 7 | 30 | 91.49 | 91.31 | 90.76 |
| 8 | 35 | 91.51 | 91.33 | 90.79 |
| 9 | 40 | 91.55 | 91.36 | 90.82 |

### *3.3 Transmittance of lens assembly with different metasurfaces*

Usually, in the LWIR region, very few transparent materials are available for imaging systems. Additionally, multilayer ARCs may suffer from thermal instability over a longer duration of use.

This motivated us to explore the possibilities of designing and fabricating ARCs out of the host/substrate material itself. The radii of curvature of the monocentric lenses are large, so they appear effectively flat over the metasurface's periodicity. In the following, we propose a design of a metasurface that acts as an ARC on a flat substrate, as shown in Fig. 11.

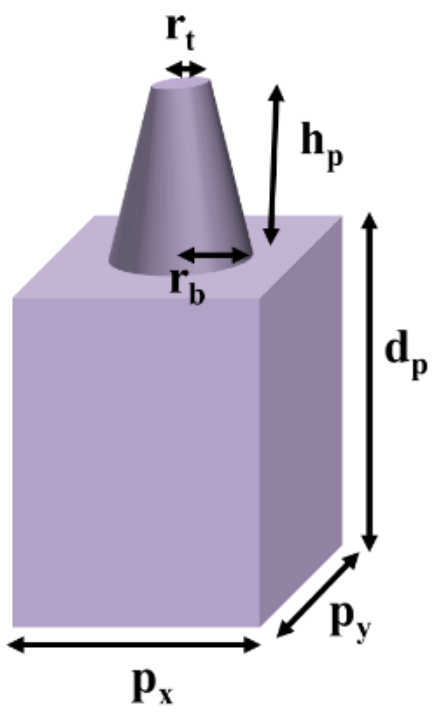

**Fig. 11:** An ARC design based on cone shape meta-structures.

In this design, we have considered $p_x = p_y = 2$ µm, $d_p = 6$ µm, and the rest of the parameters are optimized for maximum transmittance at normal incidence. For our calculation, we have chosen $r_b = 0.99$ µm, $r_t = 100$ nm, and $h_p = 3.44$ µm for the air to Ge interface. For the interface of air to Ge, the transmittances are shown in Fig. 12. From the figure, it can be seen that the transmittance remains more than 95% up to 30° angle for both TE and TM modes. The increase in transmittance is attributed to the impedance matching between the metasurface and free space, which suppresses the reflection [29, 30].

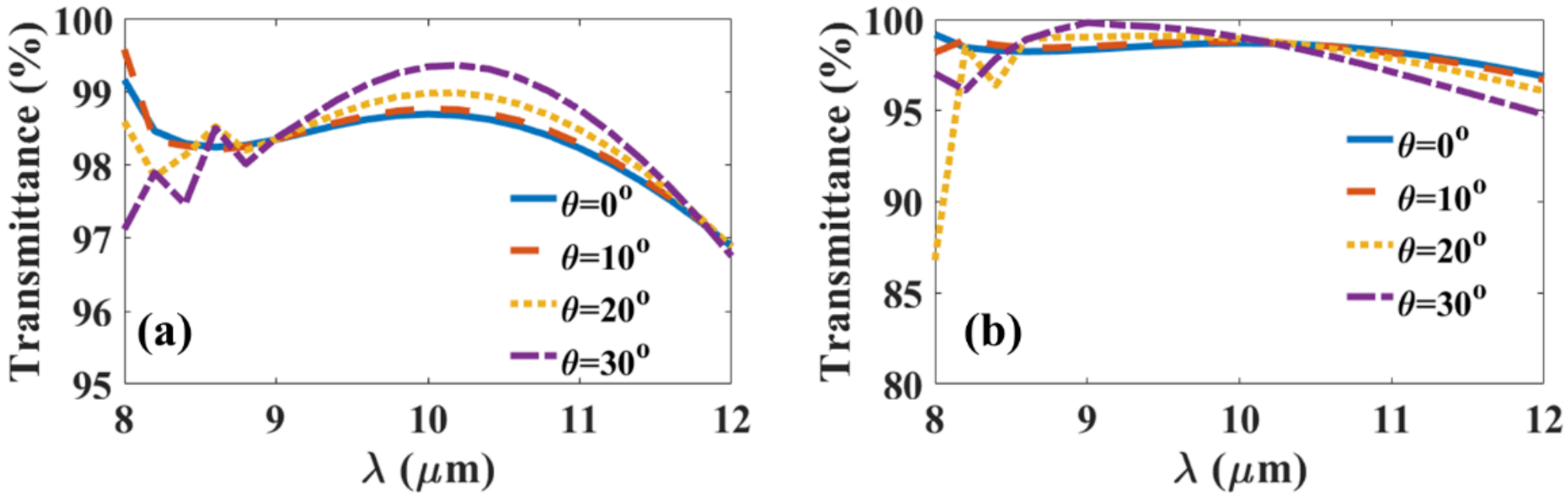

**Fig. 12:** Transmittance plots, for air to Ge interface for wider incidences (a) TE mode, and (b) TM mode.

Similarly, for air to IG6 interface, the optimized parameters are $r_p = 0.99$ µm, $w_p = 100$ nm, and $h_r = 3.94$ µm. The interface of air to IG6 also shows similar trends as reported in Fig. 13. The transmittances for both modes are more than 96% for wide field of view.

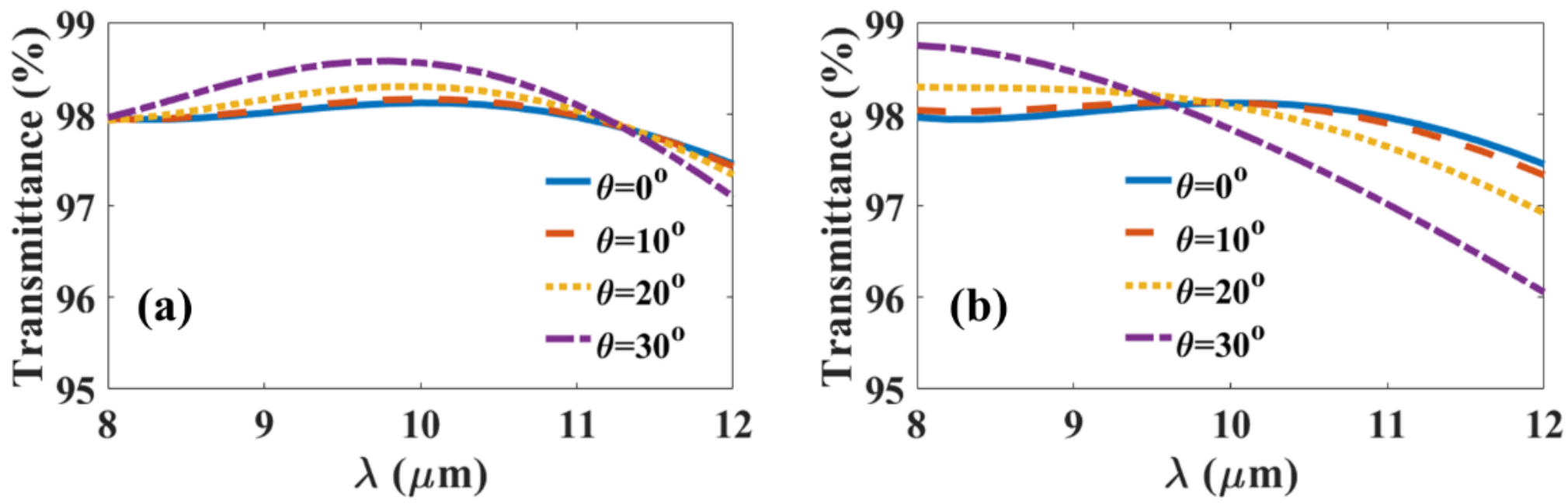

**Fig. 13:** Transmittance plots for air to IG6 interface for wider incidences (a) TE mode, and (b) TM mode.

Additionally, we also study the effect of these microstructures on Ge-IG6 interface with airgaps $d = 1\ \mu m, 2\ \mu m, 3\ \mu m, 4\ \mu m$ and $5\ \mu m$ at normal incidence. The schematic of the structure is given in Fig. 14 (a). Fig. 14 (b) illustrates that the transmittance is almost always larger than 90% over the entire wavelength range for different values of air-gaps.

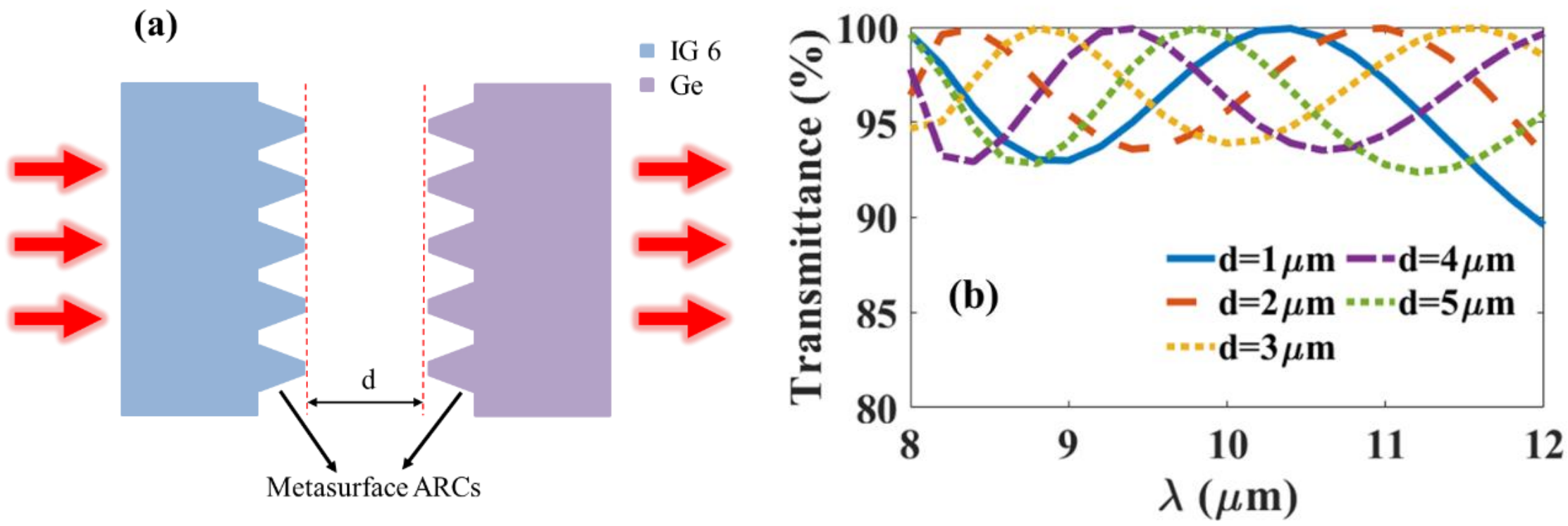

**Fig. 14:** (a) Schematic of the two lenses with metasurface ARCs and airgap, and (b) transmittance plots for IG6 to Ge interface for different widths of the airgap.

These studies establish that metasurface-based ARC has the potential to replace conventional ARCs. In addition, these microstructures can be grown on the lens surface with the lens material itself, which makes them more robust against environmental fluctuations.

## Conclusion

In this work, we studied the transmittance of IR radiation through a monocentric lens assembly, considering 97% transmitting ARCs on all surfaces. The airgap between ARCs was optimized to maximize the transmittance with variations in temperature, wavelength, and incident angles. With $10\ \mu m$ airgap, lens assembly shows mechanical stability with temperature. On increasing temperature, the airgap was found to decrease due to the expansion of the IG6 core and Ge shells. As the incident angles increase, the transmittance decreases, but for temperatures above 30°C the transmittance followed an increasing trend due to increased effective optical length for wider incident angles. For normal incidence, the average transmittance was calculated to be 49.5 %. The average transmittance over all wavelengths and all angles peaked to 37% but it increased sharply to ~45% at 40°C. The temperature variation also affects the focal spot sizes and focal lengths due to varying refractive indices of IR glasses. Importantly, with the variation in temperature, incidence angle, and wavelength, the average transmittance, focal length, and spot size (*rms*) change. In addition, we have proposed a robust alternative to conventional ARCs as meta-structures that exhibited transmittance ~ 90% through the Ge-IG6 with different airgaps interface. With the optimized shape and sizes of the metasurface, even higher transmittance can be achieved for IR radiation for a wide field-of-view.

**Acknowledgments:** This research is supported by CARS project provided by IRDE, Dehradun, India. Manish Kala wishes to acknowledge Council for Scientific and Industrial Research (CSIR), India for senior research fellowship. Pawan Singh thanks Indian Institute of Technology Roorkee, India for financial support.

**Conflict of Interest:** Authors declare no conflict of interest.

# Supplementary

# Optimization of airgap in a monocentric lens assembly and metasurface based anti-reflecting coating in the long-wave IR regime

Manish Kala[1], Pawan Singh[1], Sanjay Kumar Mishra[2], Unnikrishnan Gopinathan[2], Ajay Kumar[2], Akhilesh Kumar Mishra[1,3*]

[1]*Department of Physics, Indian Institute of Technology Roorkee, Roorkee-247667, Uttarakhand, India*
[2]*Photonics Research Lab, Instruments Research and Development Establishment, Dehradun 248008, India*
[3]*Centre for Photonics and Quantum Communication Technology, Indian Institute of Technology Roorkee, Roorkee-247667, Uttarakhand, India*
[*]*Corresponding author: akhilesh.mishra@ph.iitr.ac.in*

## S1. Variation of Spots with incident angle for different wavelengths

Here, we discuss the spatial shifts of the spot diagram on image surface at 10μm wavelength for different incident angles. Figures S1-S8 show spatial shifts & rms value of the size of spot diagram on the image plane at $0^o, 2^o, 4^o, 6^o$, and $8^o$ incident angles. Colorbars show intensity variation at these spots for different rays.

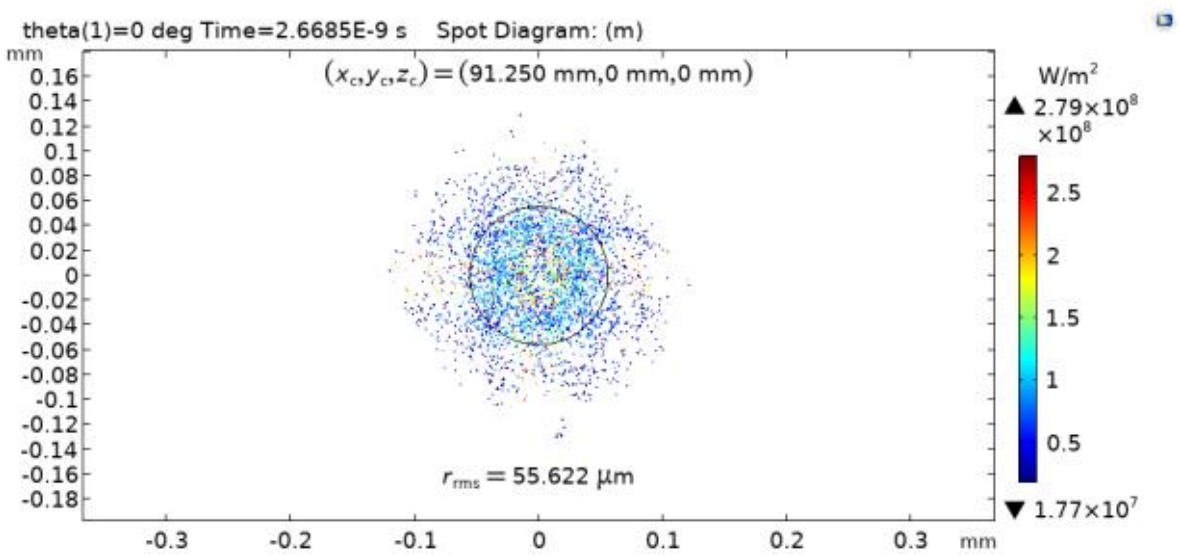

Fig. S1: Spot diagram at the image surface at $\theta = 0^o$.

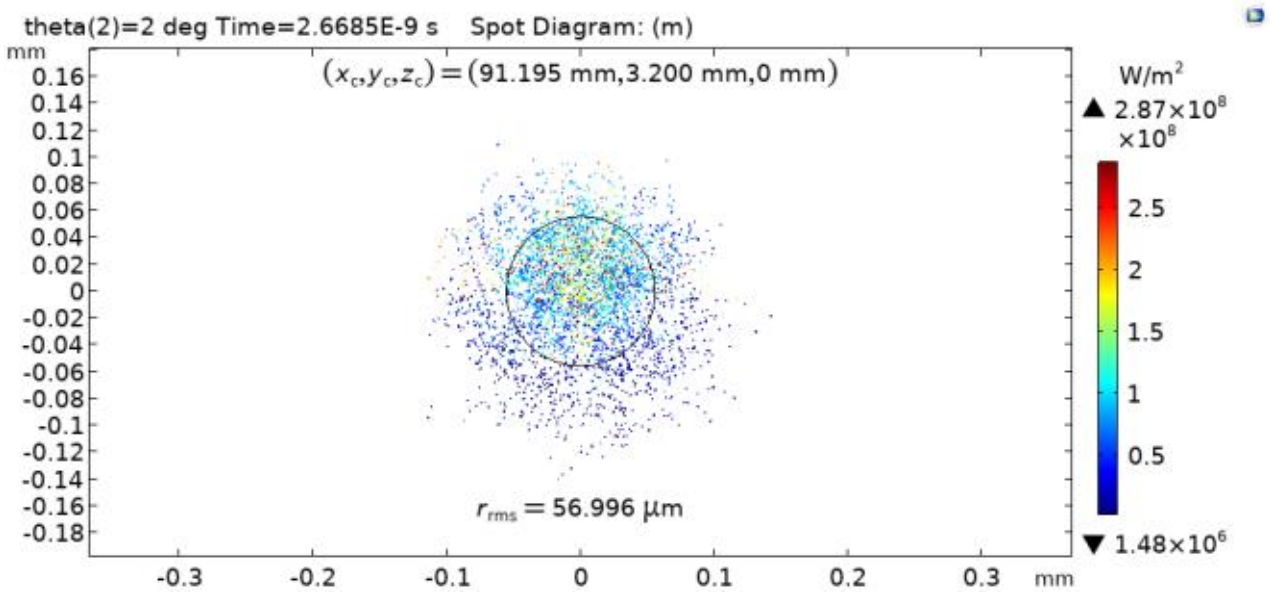

Fig. S2: Spot diagram at the image surface at $\theta = 2^o$.

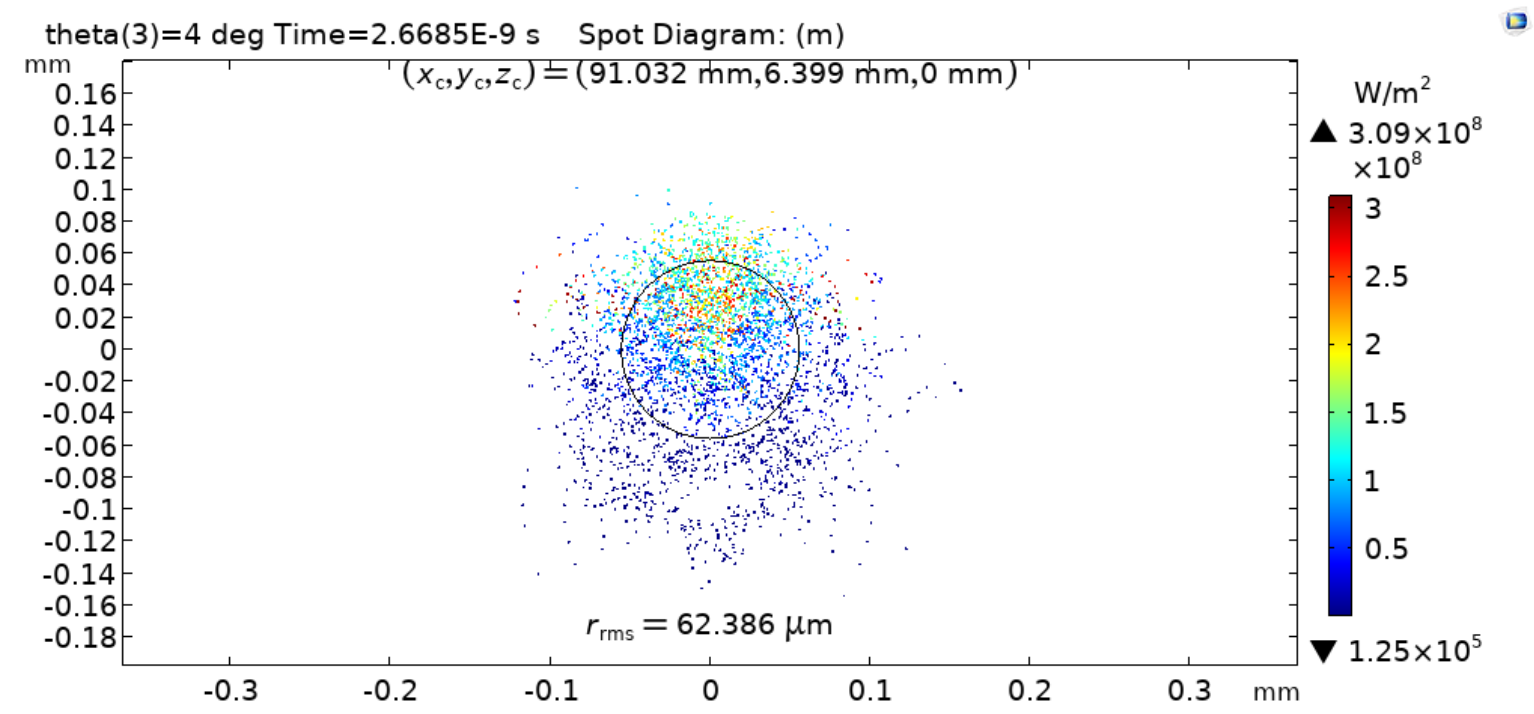

Fig. S3: Spot diagram at the image surface at $\theta = 4^o$.

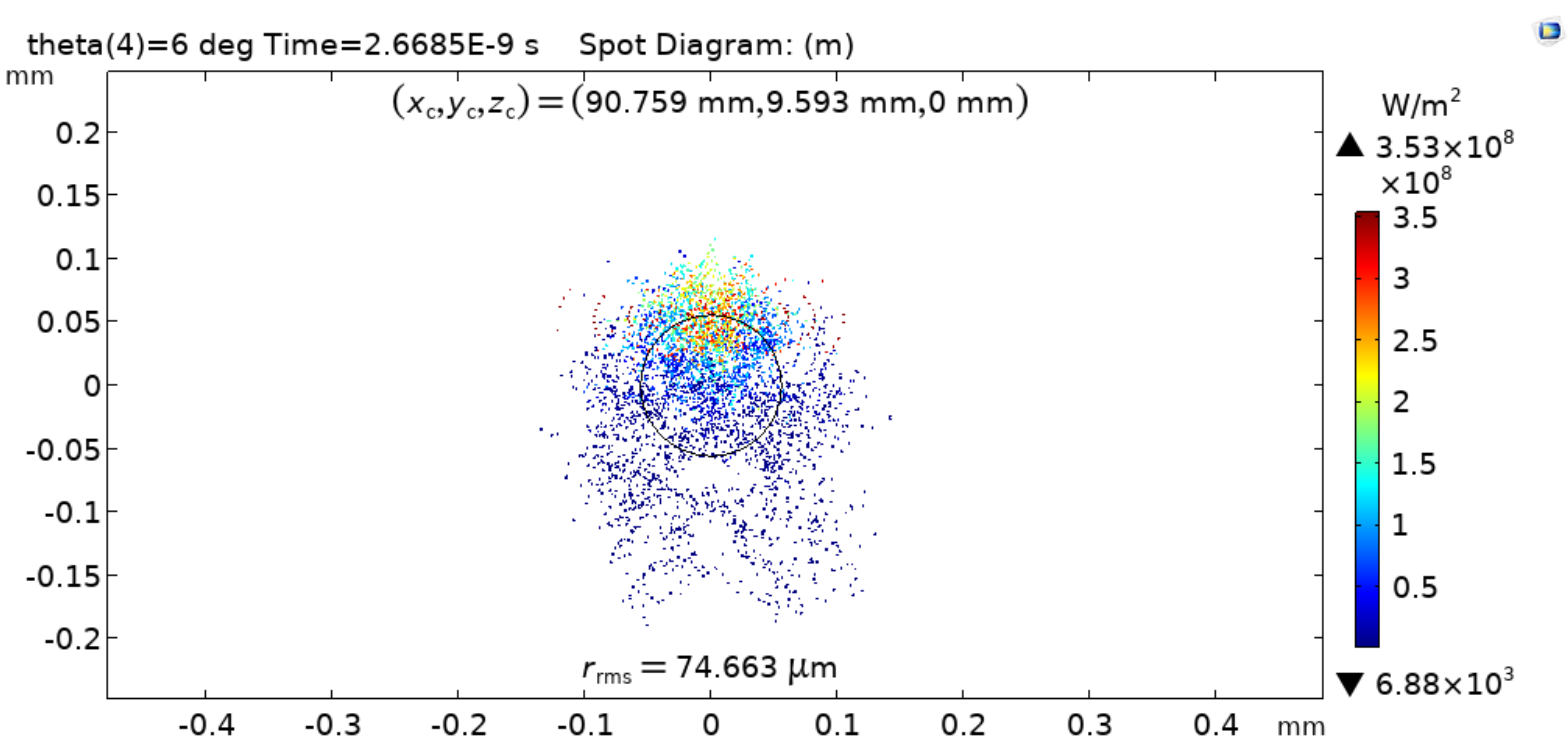

Fig. S4: Spot diagram at the image surface at $\theta = 6^o$.

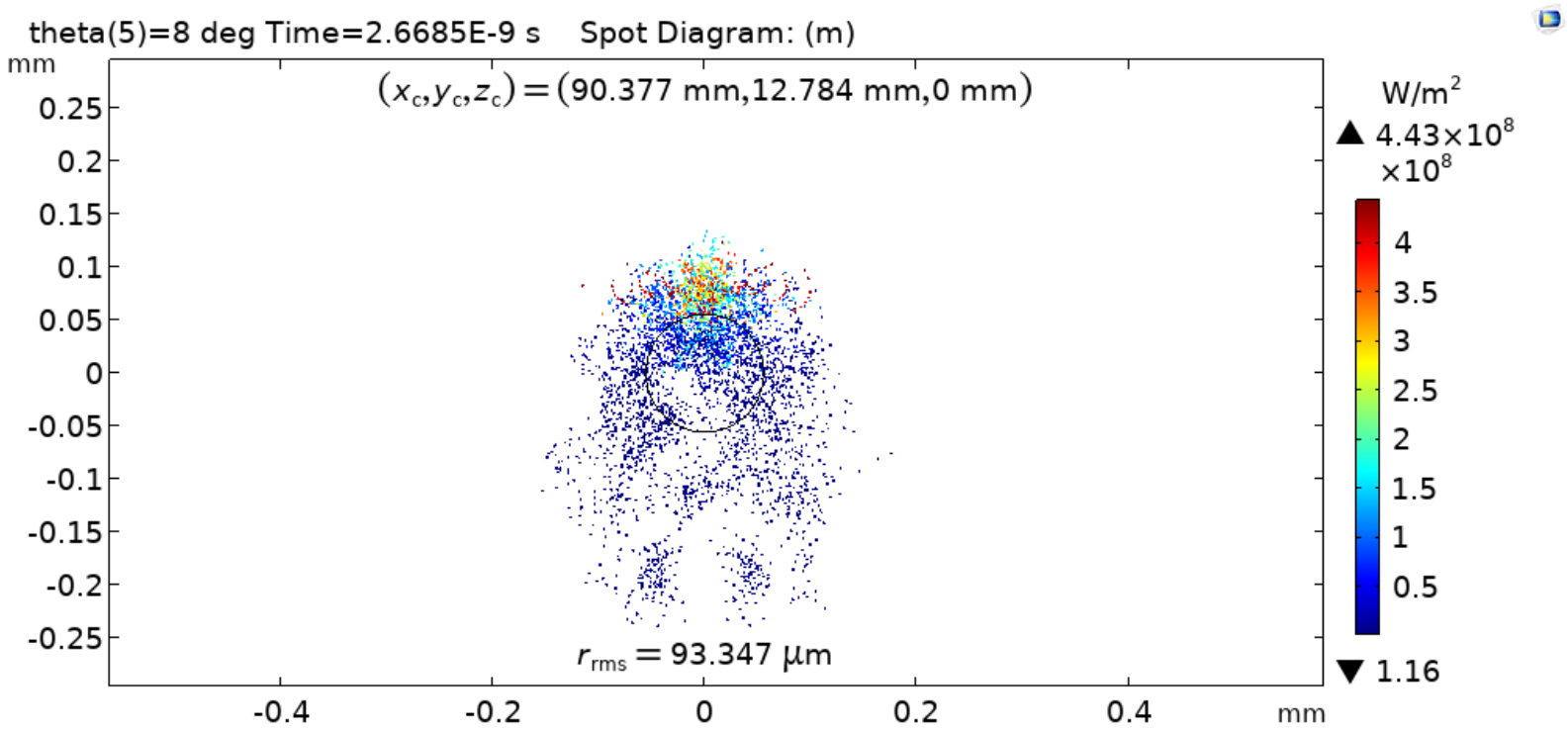

Fig. S5: Spot diagram at the image surface at $\theta = 8^o$.

## S2. Transmittance of the lens assembly for different incidence angles at different wavelengths

In this section, we study the transmittance through the lens assembly at different wavelengths for different incident angles. Figures S6-S10 depict the transmittance variation with rising temperature at different wavelengths for $0^o, 2^o, 4^o, 6^o$ and $8^o$ incident angles. From the figures, we notice that the transmittance decreases with increase in the incident angles.

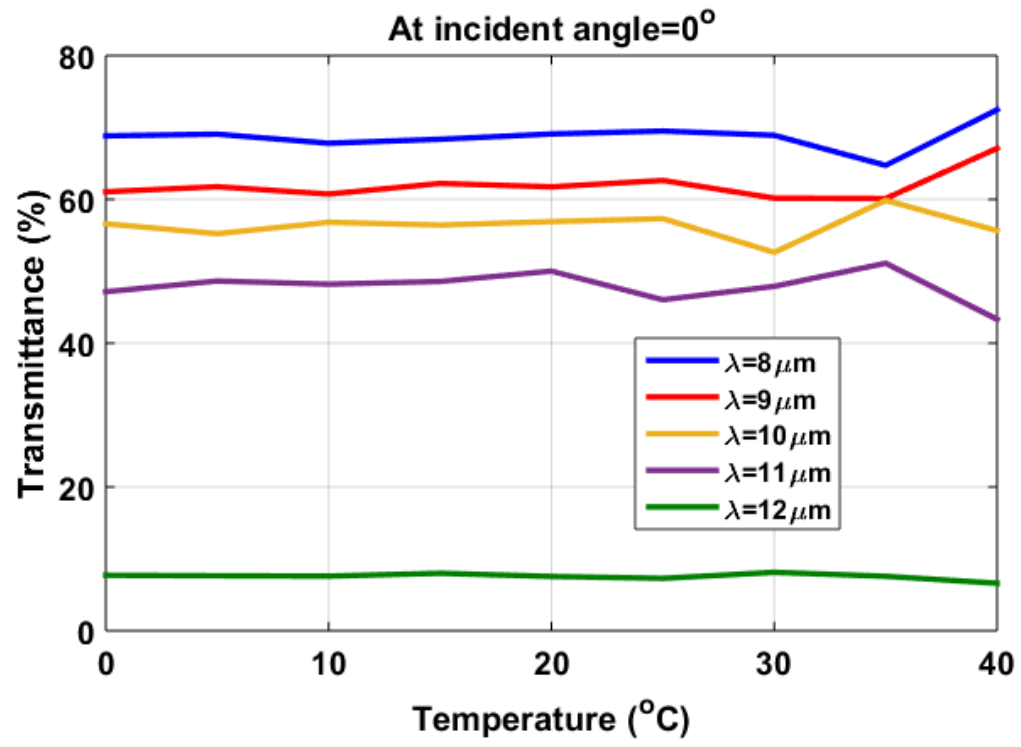

Fig. S6: Transmittance variation with temperature at $\theta = 0^o$.

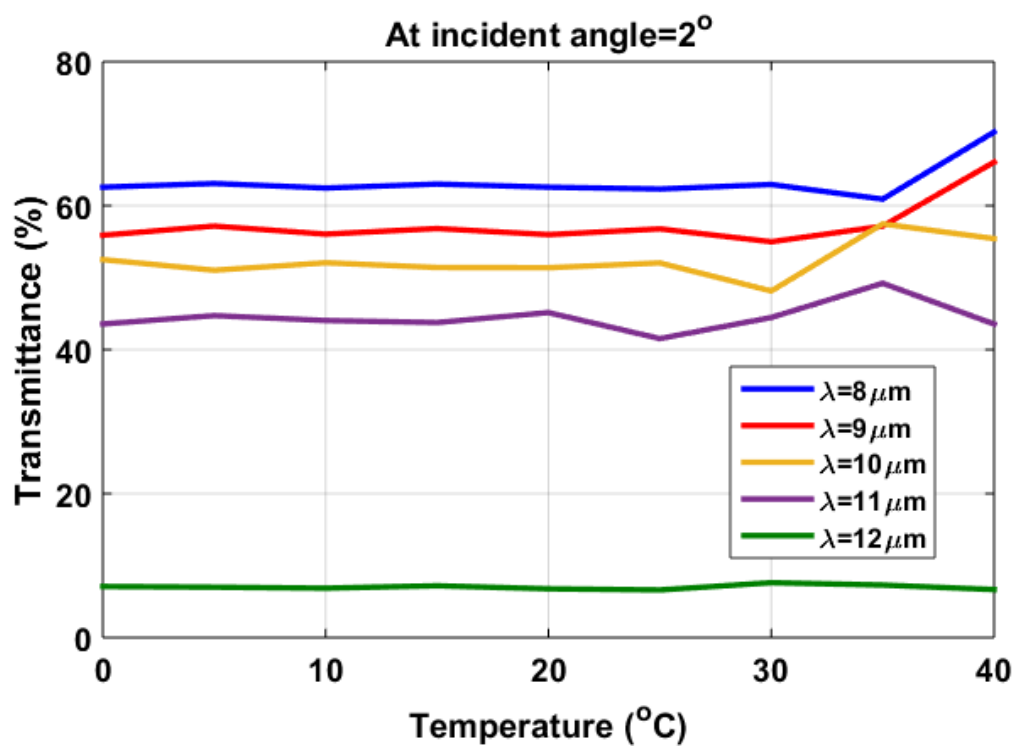

Fig. S7: Transmittance variation with temperature at $\theta = 2^o$.

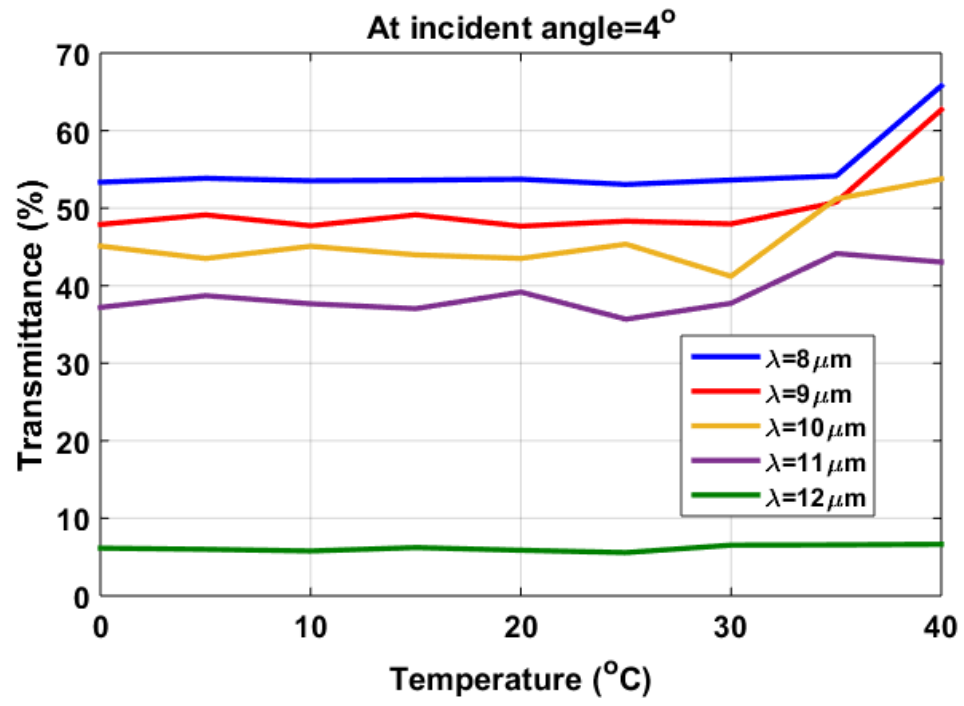

Fig. S8: Transmittance variation with temperature at $\theta = 4^o$.

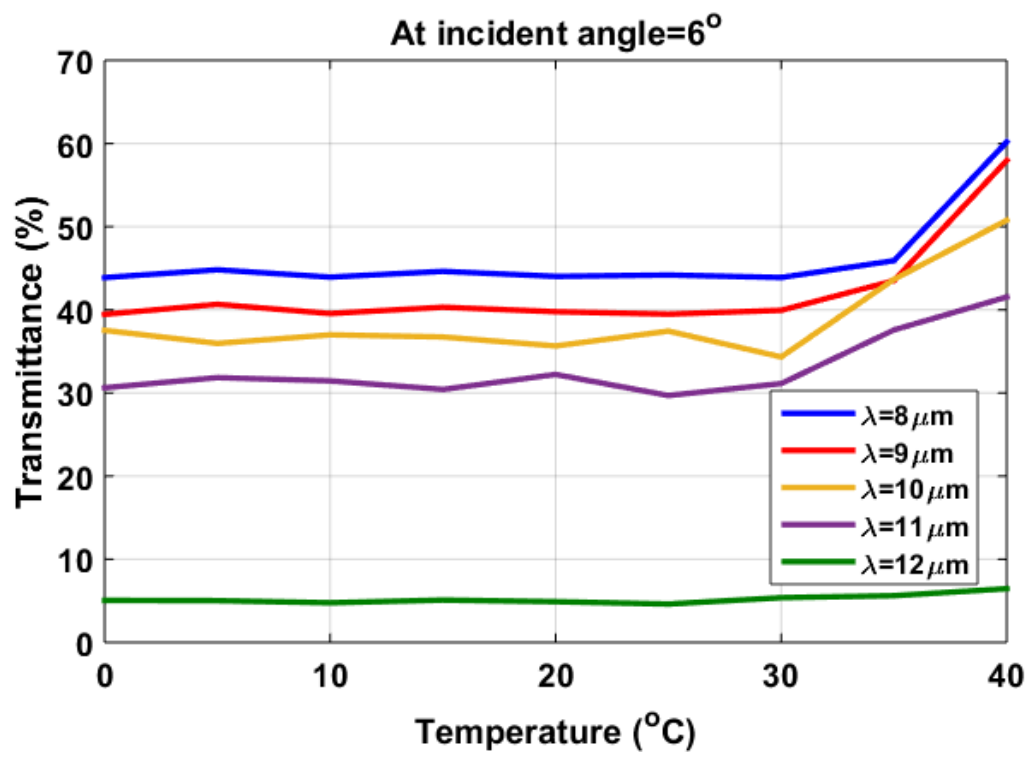

Fig. S9: Transmittance variation with temperature at $\theta = 6^o$.

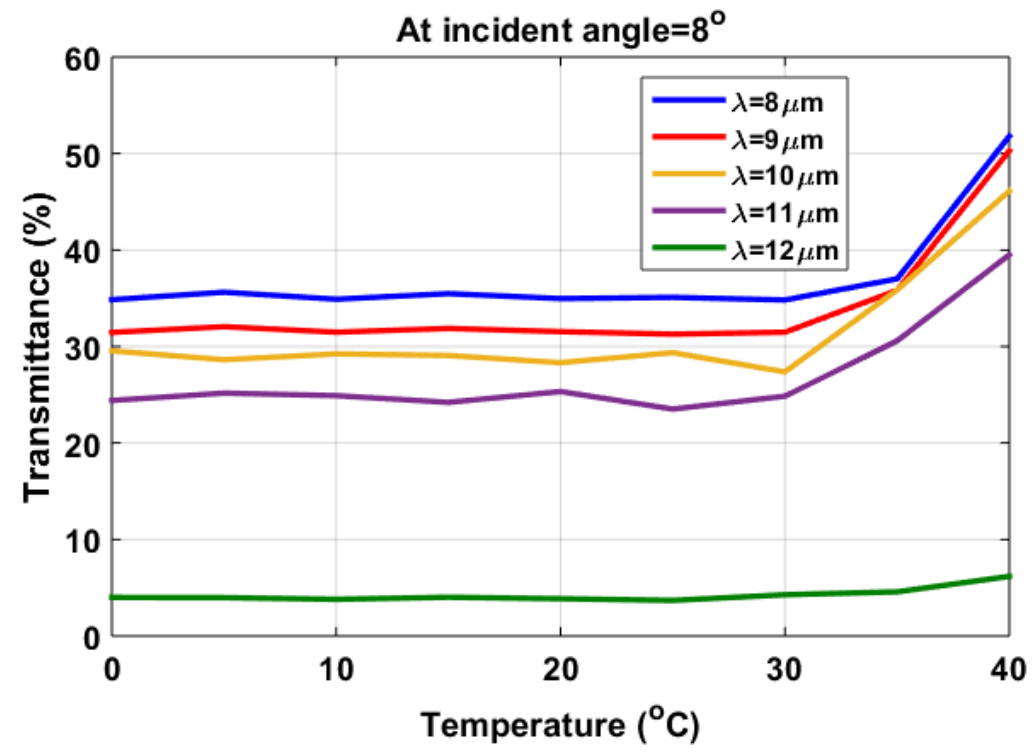

Fig. S10: Transmittance variation with temperature at $\theta = 8^o$.

## S3. Variation of the spot diagrams with temperature and incident angles

In this section, we show the variation in the spot size and its spatial shift with increasing temperature from 273K-313K (0 to 40°C). Figures S11-S19 reveal that the spot diagram for 10 μm wavelength shift on the image surface and the *rms* values of radius also change accordingly, as mentioned in the figures.

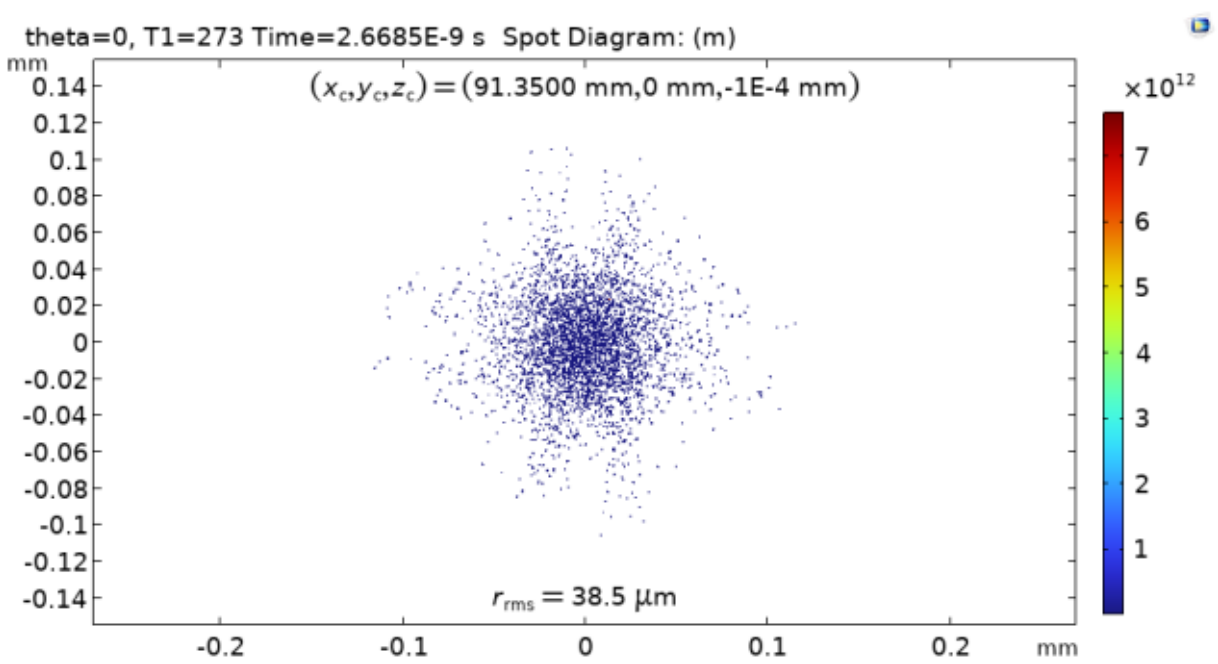

Fig. S11: Spot diagram at image surface at temperature $0^o C$ $(273\ K)$.

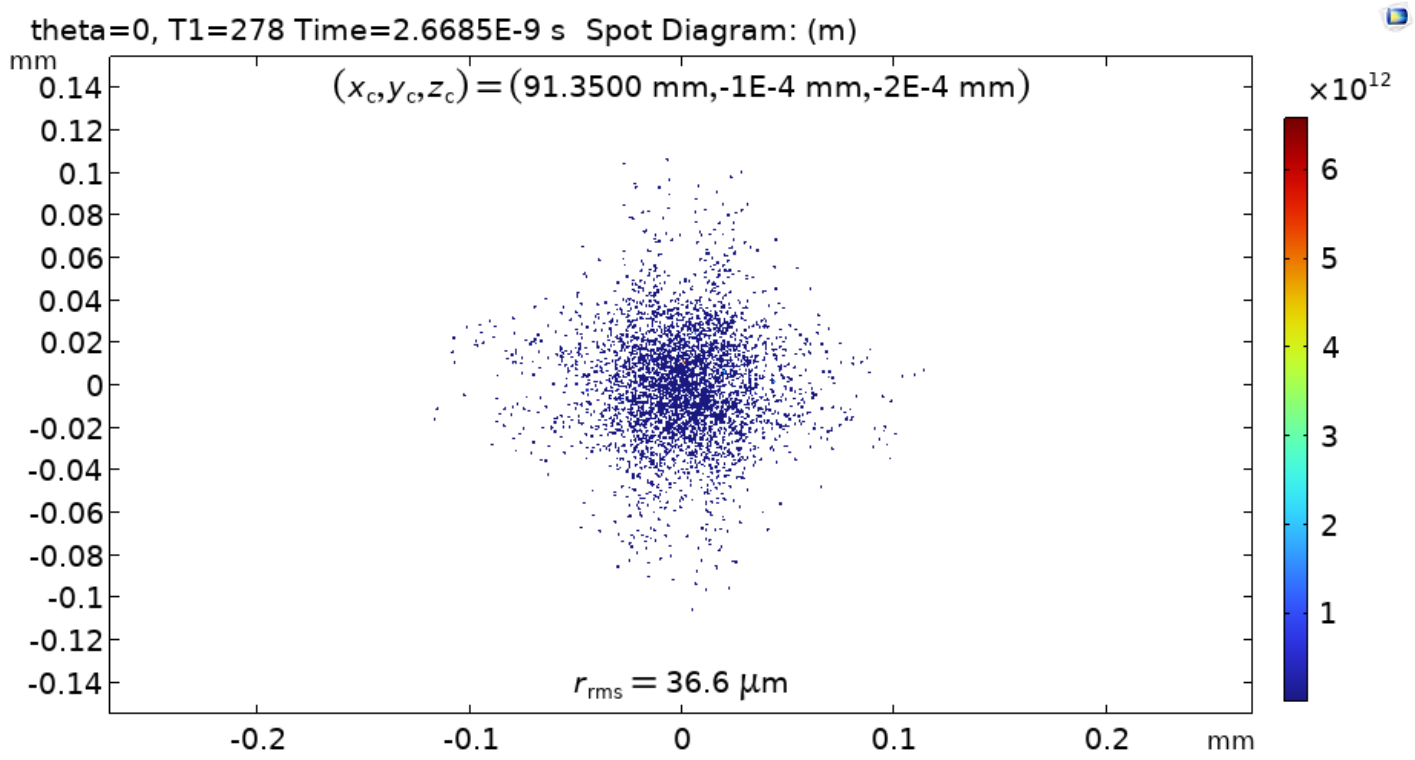

Fig. S12: Spot diagram at image surface at temperature $5^o C$ $(278\ K)$.

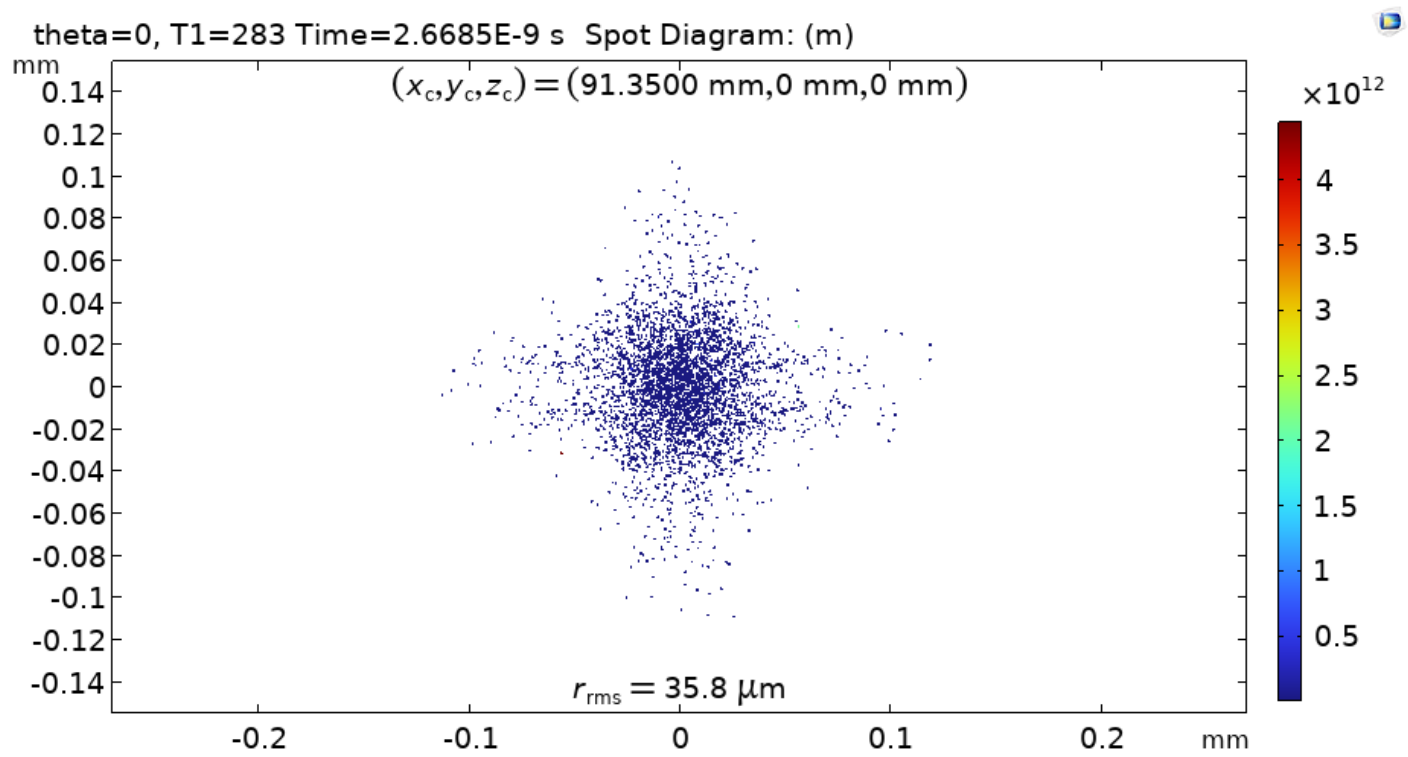

Fig. S13: Spot diagram at image surface at temperature $10^o C$ $(283\ K)$.

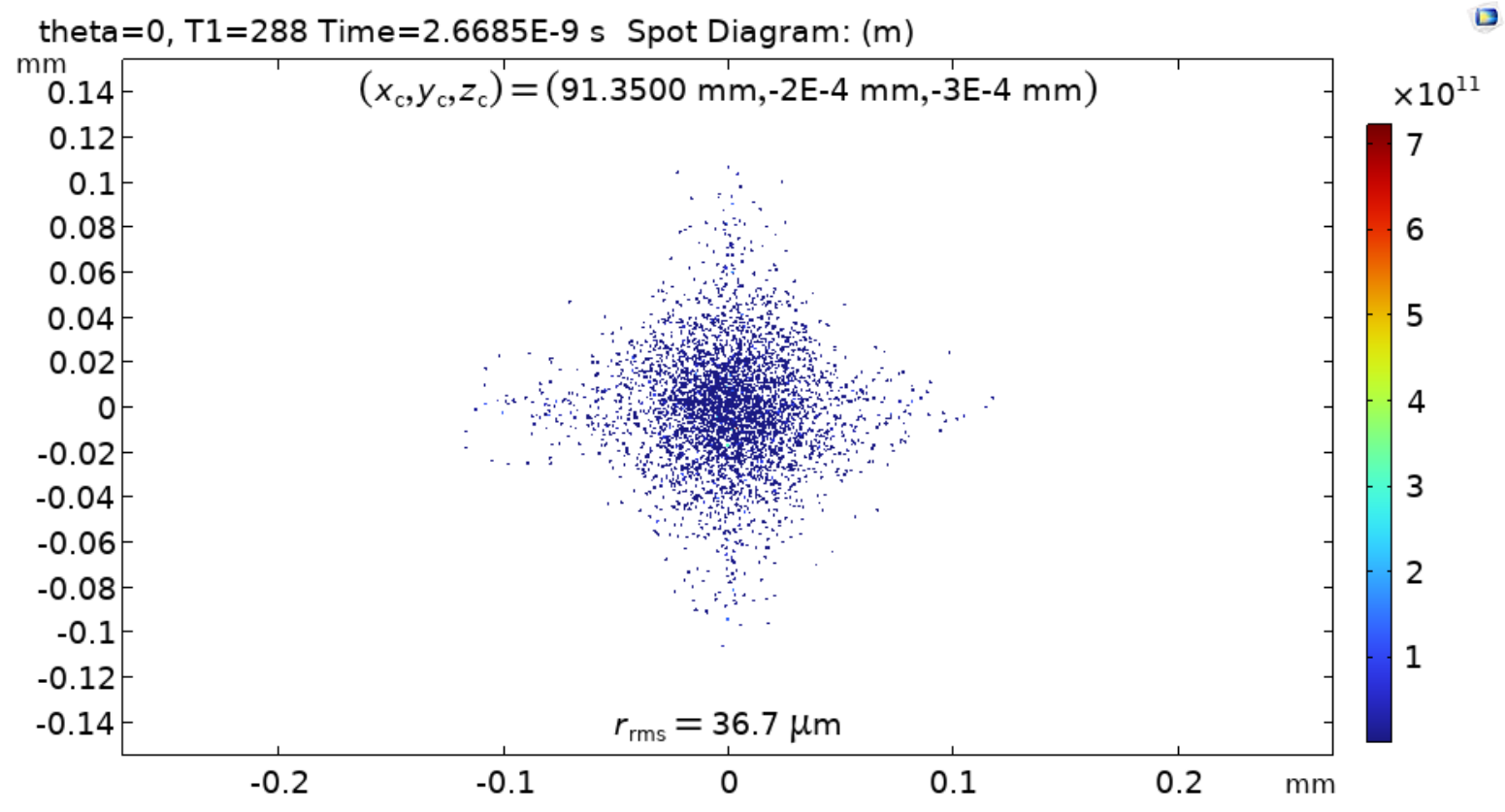

Fig. S14: Spot diagram at image surface at temperature $15^o C$ $(288\ K)$.

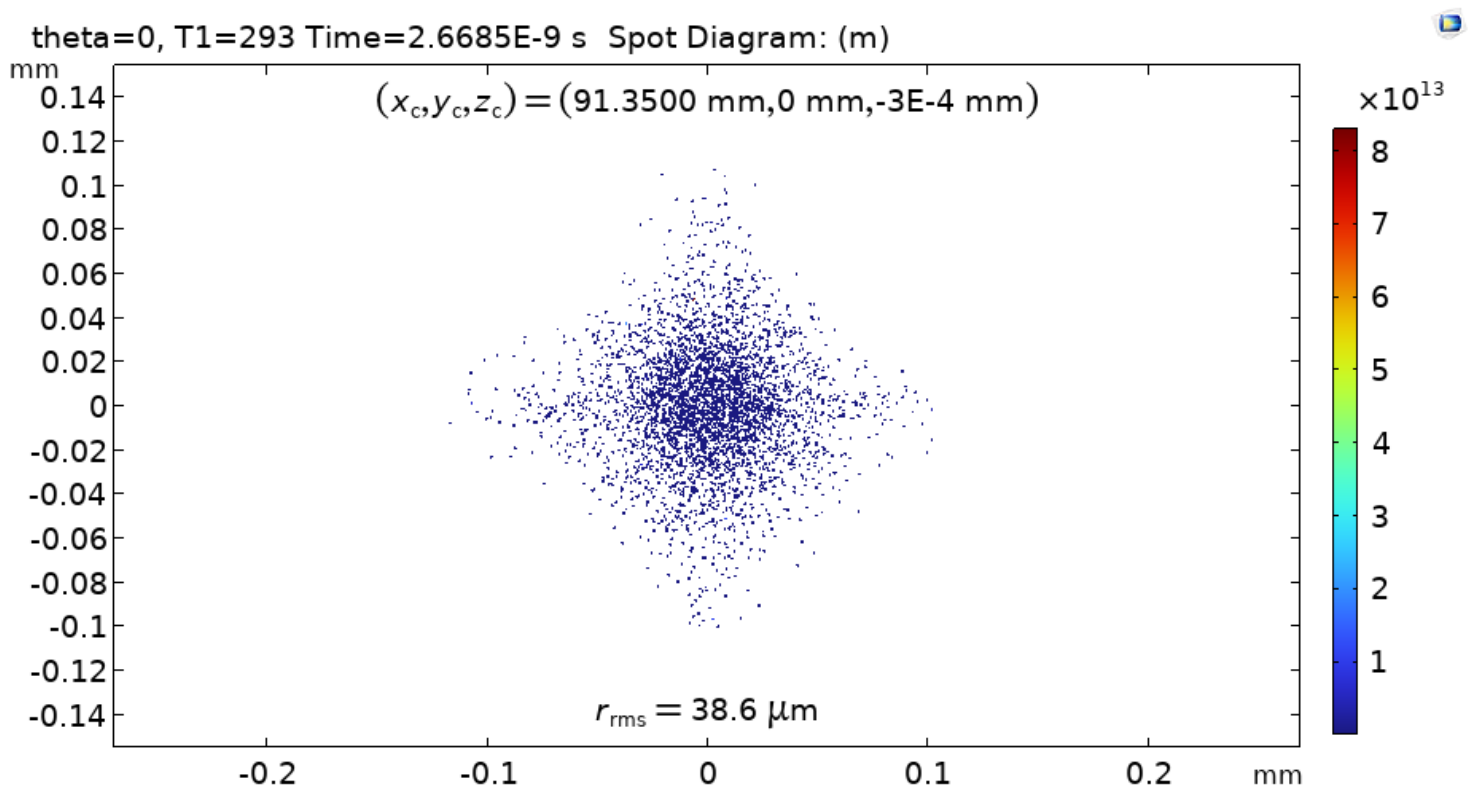

Fig. S15: Spot diagram at image surface at temperature $20^{o}C$ $(293K)$.

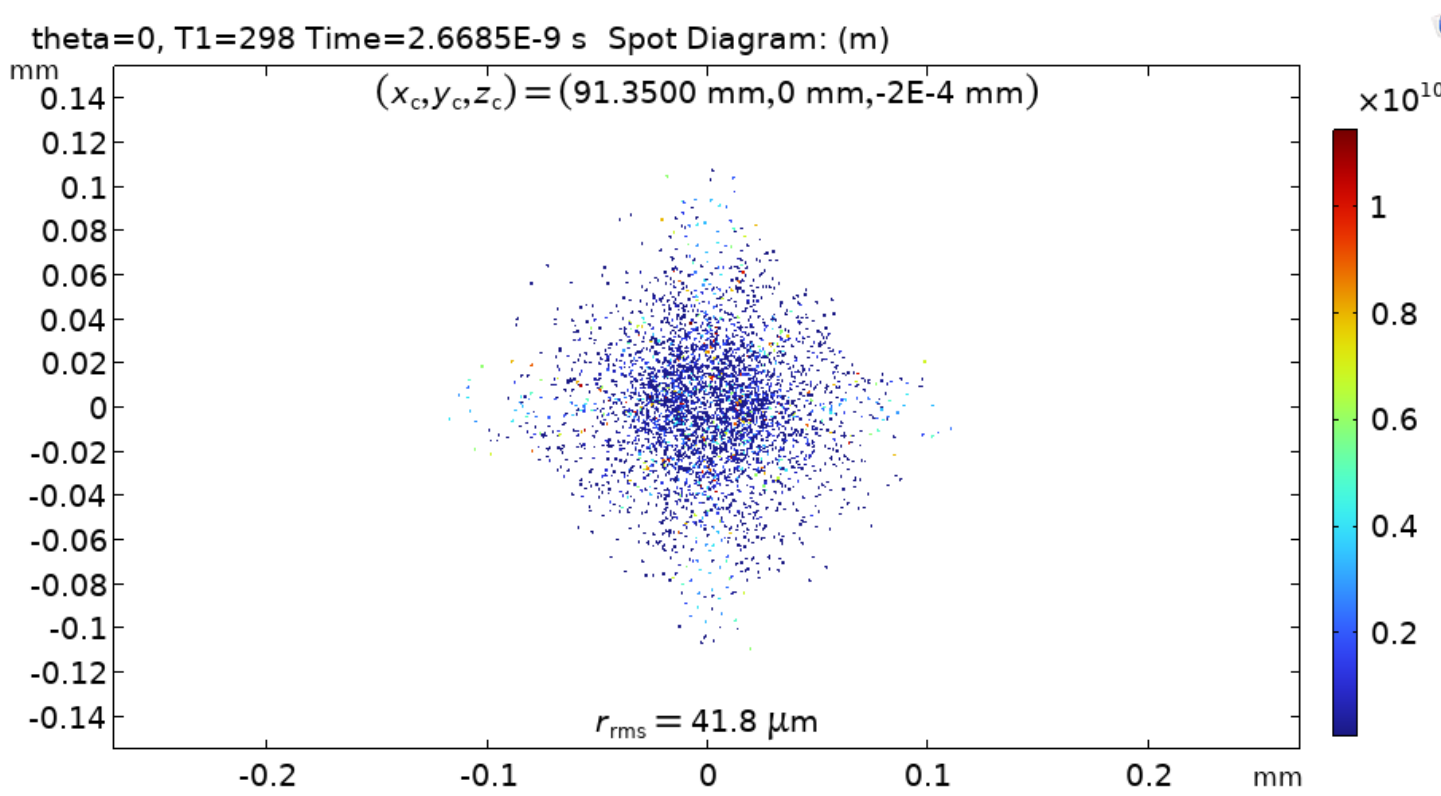

Fig. S16: Spot diagram at image surface at temperature $25^{o}C$ $(293\ K)$.

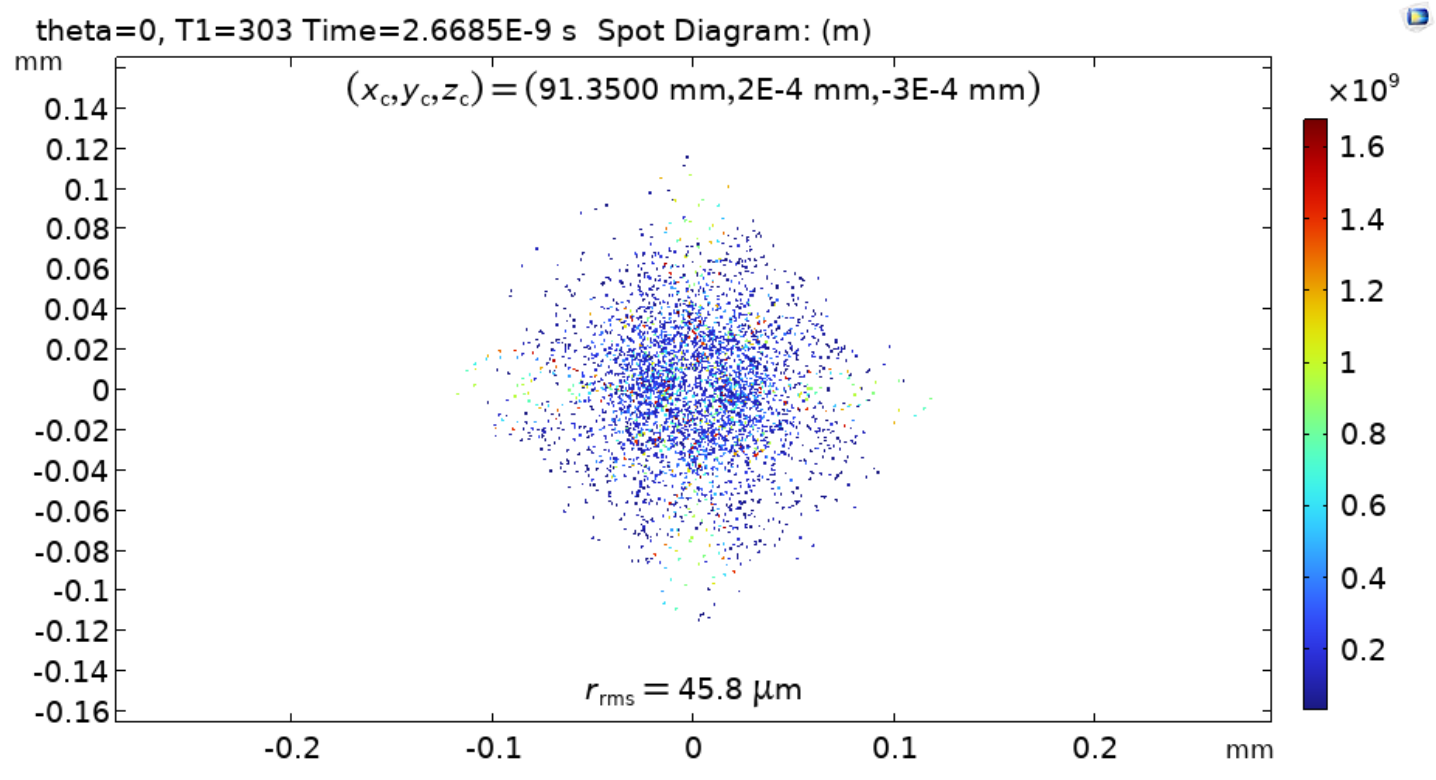

Fig. S17: Spot diagram at image surface at temperature $30^{o}$ $(303\ K)$.

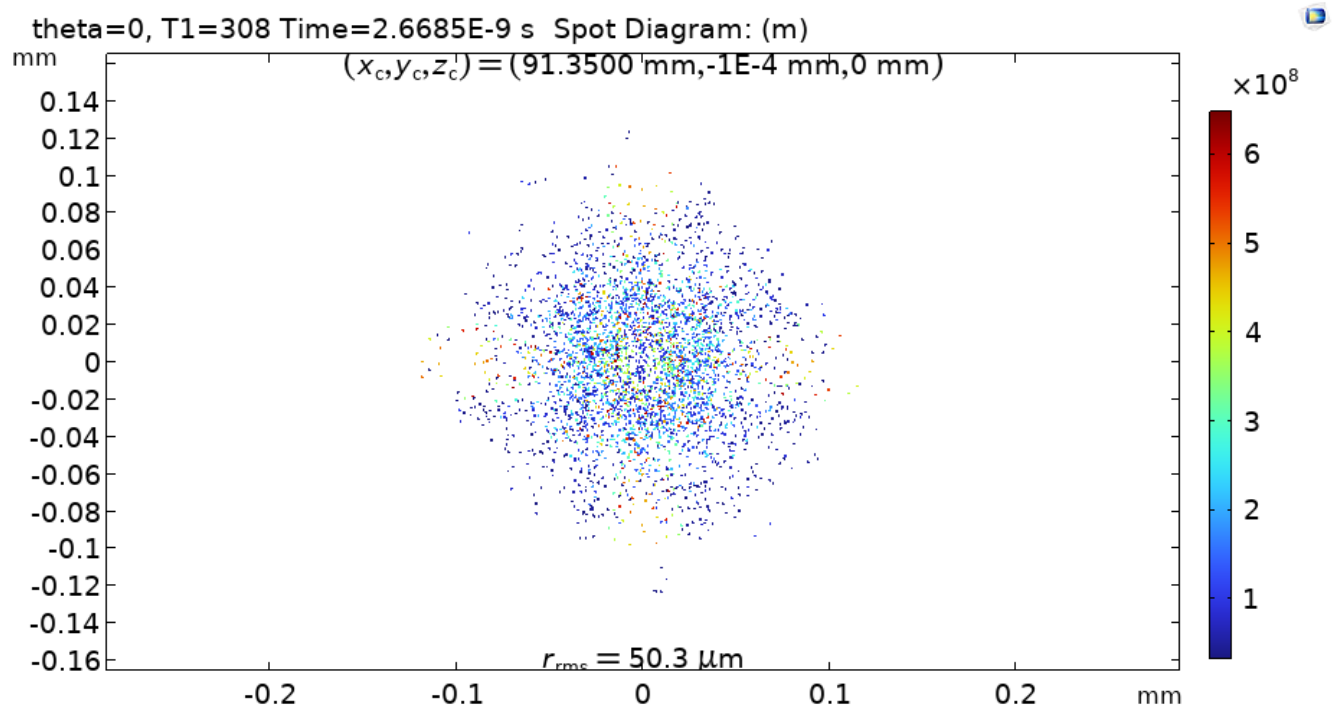

Fig. S18: Spot diagram at image surface at temperature $35^o(308\ K)$.

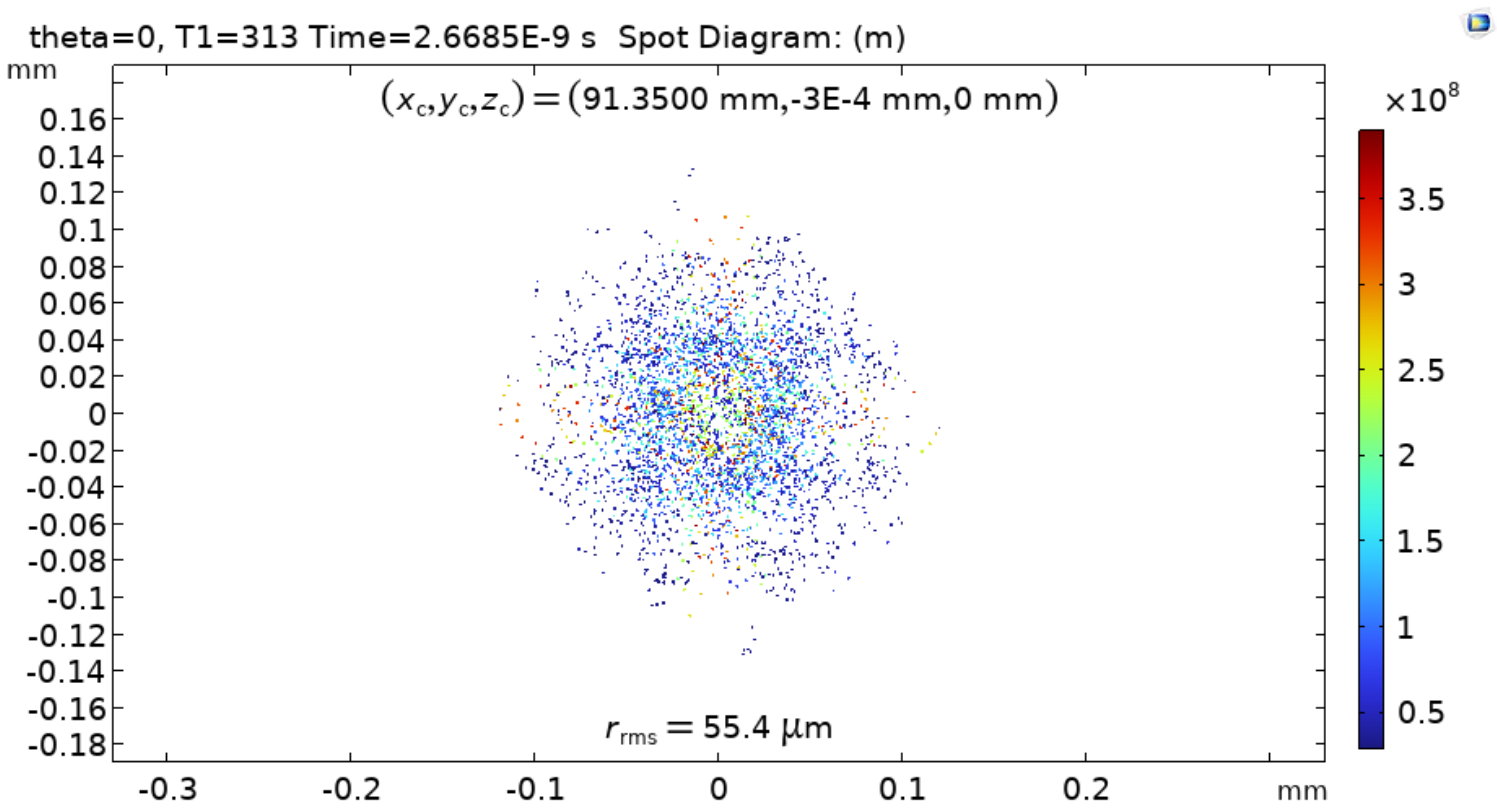

Fig. S19: Spot diagram at image surface at temperature $40^o$ $(313\ K)$.

## S4. Variation of the focal length of the monocentric lens assembly with temperature and incident angles

In this section, we discuss the variation in focal length of the lens assembly with temperature and incidence angle. With an increase in temperature, the focal length increases, while it decreases with incident angles as reported in Table S1.

Table S1: Variation in focal length with temperature and incident angles.

| S. No. | $T$ $(^oC)$ | Calculated focal length (mm) at 10 μm wavelength for different incident angles $(\theta)$ | | | | | | | | |
|---|---|---|---|---|---|---|---|---|---|---|
| | | $0^o$ | $1^o$ | $2^o$ | $3^o$ | $4^o$ | $5^o$ | $6^o$ | $7^o$ | $8^o$ |
| 1 | 0 | 91.287 | 91.266 | 91.241 | 91.187 | 91.089 | 91.001 | 90.884 | 90.739 | 90.571 |
| 2 | 5 | 91.316 | 91.302 | 91.276 | 91.219 | 91.154 | 91.032 | 90.918 | 90772 | 90.604 |
| 3 | 10 | 91.344 | 91.335 | 91.303 | 91.256 | 91.178 | 91.065 | 90.952 | 90.802 | 90.636 |
| 4 | 15 | 91.373 | 91.365 | 91.335 | 91.286 | 91.215 | 91.098 | 90.992 | 90.833 | 90.667 |
| 5 | 20 | 91.407 | 91.399 | 91.376 | 91.323 | 91.246 | 91.131 | 91.016 | 90.866 | 90.699 |
| 6 | 25 | 91.440 | 91.427 | 91.403 | 91.356 | 91.280 | 91.167 | 91.048 | 90.901 | 90.731 |
| 7 | 30 | 91.497 | 91.465 | 91.403 | 91.385 | 91.312 | 91.201 | 91.078 | 90.931 | 90.764 |
| 8 | 35 | 91.512 | 91.502 | 91.470 | 91.418 | 91.338 | 91.233 | 91.110 | 90.965 | 90.795 |
| 9 | 40 | 91.556 | 91.535 | 91.503 | 91.454 | 91.369 | 91.265 | 91.142 | 90.998 | 90.827 |